\documentclass{article}
\usepackage{amsfonts}
\usepackage{amsmath}
\usepackage{eucal}
\usepackage{amssymb}
\usepackage{graphicx}
\usepackage{graphpap}  
\usepackage{epic, xypic} 
\newtheorem{theorem}{Theorem}[section]
\newcommand{\sq}{\sqrt{3}}
\newcommand{\di}[1]{\frac{#1}{\sqrt{3}}}
\renewcommand{\theequation}{\thesection.\arabic{equation}}
\newcommand{\cvd}{\begin{flushright}$\Box$\end{flushright}}
\newcommand{\ad}{{\rm ad}\;}
\newcommand{\dd}{{\rm d}}
\newcommand{\ee}{{\rm e}}
\newcommand{\tr}{{\rm Tr}\;}
\newcommand{\sgn}{{\rm sgn}\,}
\newcommand{\rif}[1]{(\ref{#1})}
\newcommand{\der}[2]{{\partial #1\over\partial #2}}
\newcommand{\bra}[1]{\left< #1\right|}
\newcommand{\ket}[1]{\left| #1\right>}
\newcommand{\sign}{{\mathrm{sign}}}
\newcommand{\eq}{\begin{equation}}
\newcommand{\feq}{\end{equation}}
\newcommand{\eqn}{\begin{eqnarray}}
\newcommand{\feqn}{\end{eqnarray}}
\newcommand{\arr}{\begin{eqnarray*}}
\newcommand{\farr}{\end{eqnarray*}}
\newcommand{\bea}{\begin{array}}
\newcommand{\ea}{\end{array}}
\newcommand{\dem}{{\frac 12}}
\newcommand{\inv}[1]{{1\over #1}}
\newcommand{\intS}{\int_\Sigma}
\newcommand{\intM}{\int_{\cal M}}
\newcommand{\intdM}{\int_{\partial\cal M}}
\newcommand{\dM}{{\partial\cal M}}
\newcommand{\M}{{\cal M}}
\newcommand{\RR}{{\cal R}}
\newcommand{\DD}{{\cal D}}
\newcommand{\HH}{{\cal H}}
\newcommand{\VV}{{\cal V}}
\newcommand{\lie}{{\cal L}}
\newcommand{\G}{{\cal G}}
\newcommand{\p}{\partial}
\newcommand{\w}{\wedge}
\newcommand{\rerg}{r_{\rm erg}}
\newcommand{\ig}{{g^{-1}}}
\newcommand{\ih}{{h^{-1}}}
\newcommand{\At}{{\tilde A}}
\newcommand{\lp}{\left(}
\newcommand{\rp}{\right)}
\newcommand{\de}{\partial}
\newcommand{\ld}{\ldots}
\newcommand{\al}{\alpha}
\newcommand{\tc}{\tilde c}
\newcommand{\cosech}{{\mathrm{cosech}}}
\font\mybb=msbm10 at 12pt
\def\bb#1{\hbox{\mybb#1}}
\def\bZ {\bb{Z}}
\def\bR {\bb{R}}
\def\bE {\bb{E}}
\def\bT {\bb{T}}
\def\bM {\bb{M}}
\def\bC {\bb{C}}
\def\bO {\bb{O}}
\newcommand{\cG}{{\cal G}}
\newcommand{\cHH}{{\cal H}}
\newcommand{\cP}{{\cal P}}

\newcommand{\SSu}{\rm Sum}
\newcommand{\Coe}{\rm Coefficient}
\newcommand{\Part}{\rm Part}
\newcommand{\QQ}{\rm QQ}
\newcommand{\Qm}{\rm Qm}
\newcommand{\QP}{\rm QP}
\newcommand{\Conj}{\rm Conj}
\newcommand{\Octp}{\rm OctP}
\newcommand{\Octps}{\rm OctPS}
\newcommand{\Append}{\rm Append}
\newcommand{\Do}{\rm Do}
\newcommand{\Array}{\rm Array}
\newcommand{\iem}{\frac {i}{2}}

\numberwithin{equation}{section}

\begin{document}
\begin{titlepage}
\begin{flushright}
UCB-PTH-07/14\\ IFIC/07$-$41\\ FTUV/07$-$MMDD\\
\end{flushright}
\vskip 7mm
\begin{center}
\renewcommand{\thefootnote}{\fnsymbol{footnote}}
{\Large \bf Mapping the geometry of the \boldmath{$E_6$} group}

\vskip 7mm {\large \bf { Fabio
Bernardoni$^{1}$\footnote{Fabio.Bernardoni@ific.uv.es}, Sergio
L.~Cacciatori$^{2}$\footnote{sergio.cacciatori@uninsubria.it},
Bianca L.~Cerchiai$^{3}$\footnote{BLCerchiai@lbl.gov} and
Antonio Scotti$^{4}$\footnote{antonio.scotti@gmail.com}}}\\
\renewcommand{\thefootnote}{\arabic{footnote}}
\setcounter{footnote}{0} \vskip 10mm {\small $^1$Departament de
F\'\i sica Te\`orica, IFIC, Universitat de Val\`encia - CSIC
\\ Apt. Correus 22085, E-46071 Val\`encia, Spain. \\

\vspace*{0.5cm}

$^2$ Dipartimento di Scienze Fisiche e Matematiche, \\
\hspace*{0.15cm} Universit\`a dell'Insubria, \\
\hspace*{0.15cm} Via Valleggio 11, I-22100 Como. \\
\vspace*{0.5cm}

$^3$ Lawrence Berkeley National Laboratory\\
Theory Group, Bldg 50A5104\\
1 Cyclotron Rd, Berkeley CA 94720 USA\\
\vspace*{0.5cm}

$^4$
Dipartimento di Matematica dell'Universit\`a di Milano,\\
Via Saldini 50, I-20133 Milano, Italy. \\

\vspace*{0.5cm}

$^5$ INFN, Sezione di Milano, Via Celoria 16, I-20133 Milano.

}
\end{center}
\vspace{0.4cm}
\begin{center}
{\bf Abstract}
\end{center}
{\small { In this paper we present a construction for the compact
form of the exceptional Lie group $E_6$ by exponentiating the
corresponding Lie algebra $\mathfrak {e}_6$, which we realize as the
the sum of $\mathfrak {f}_4$, the derivations of the exceptional Jordan
algebra $J_3$ of dimension 3 with octonionic entries, and
the right multiplication by the elements of $J_3$ with vanishing trace.
Our parametrization is a generalization of the Euler angles
for $SU(2)$ and it is based on the fibration of $E_6$ via a $F_4$ subgroup as the fiber.
It makes use of a similar construction we have performed in a
previous article for $F_4$. An interesting first application of these
results lies in the fact that we are able to determine an explicit expression
for the Haar invariant measure on the $E_6$ group manifold.}}

\end{titlepage}

\section{Introduction.}
The Standard Model (SM) provides a very good description of elementary particle
physics. However, despite its success there are reasons to go beyond it:
for example the recent discovery of neutrino oscillations, the fine tuning
of the mixing matrices, the hierarchy problem, the difficulty in including
gravity and so on.

A starting point could be the fact that the renormalization flow of the
coupling constants suggests the unification of gauge interactions
at energies of the order of $10^{15}$ GeV, which can be improved (fine tuned)
by supersymmetry. It is natural to expect the gauge group $G$ of the GUT
theory to be a simple group. Obviously, at low energies the GUT model must
reproduce the SM physics, so that not only $G$ should contain
$SU(3)_c \times SU(2)_L \times U(1)_Y$ (the SM gauge group), but it should also
predict the correct spectra after spontaneous symmetry breaking.
The increasing accuracy in the analysis of particles spectra imposes even more
restrictions in the possible choices for $G$. E.g., it is already
known that the current estimate for the lower bound of the proton lifetime
rules out some of the GUT models candidates. Recently, the
particular structure of the neutrino mixing matrix seems to suggest that a
good candidate for a GUT could be based on the semidirect product between
the exceptional group $E_6^4$ and the discrete group $S_4$ \cite{caramor,das}.\\
In general, while the local properties of the group $G$, which for a Lie group
are all encoded in the corresponding Lie algebra, are enough to perform a
perturbative analysis, in order to obtain non perturbative results, such
as instantonic calculations or lattice simulations, a knowledge of the entire
group structure is required, in particular of the invariant measure on the
group in a suitable global parametrization. There are many ways to give such
an explicit expression for the Haar measure on a Lie group. However for large
dimensions it becomes quite hard to find a realization, which is able at
the same time to provide both a reasonably simple form for the measure and
an explicit determination for the range of the angles.
In this paper we solve this problem for the exceptional Lie group $E_6$,
by constructing a generalization of the Euler parametrization, with a
technique we have introduced in \cite{noi} and fully developed in \cite{f4}.
In section~\ref{algebra} we explain how the $\mathfrak{e}_6$
algebra can be represented using a theorem due to Chevalley and Schafer.
The construction of the group and the determination of the corresponding
Haar measure is made in section~\ref{group}.

\section{The construction of the \boldmath{$\mathfrak{e}_6$} algebra.} \label{algebra}
Our starting point for the construction of the exceptional algebra $\mathfrak{e}_6$ is a Theorem
due to Chevalley and Schafer \cite{Chevalley-schafer} which we rewrite here for convenience

\begin{theorem}
The exceptional simple Lie algebra $\mathfrak{f}_4$ of dimension $52$ and rank $4$ over $K$ is the derivation algebra
$\mathfrak {D}$ of the exceptional Jordan algebra $\mathfrak {J}$ of dimension $27$ over $K$. The exceptional simple
Lie algebra $\mathfrak{e}_6$ of dimension $78$ and rank $6$ over $K$ is the Lie algebra
\eqn
\mathfrak{D}+\{R_Y\}\ , \qquad \ Tr{Y}=0 \ ,
\feqn
spanned by the derivations of $\mathfrak{J}$ and the right multiplications of elements $Y$ of trace 0.
\end{theorem}
We will refer to \cite{fulton-harris} for the notations. Then the exceptional Jordan algebra is the algebra $J_3$.
For our purposes $K$ will be $\mathbb R$ or $\mathbb C$. The right multiplication is $R_Y (X)=Y\circ X$
and the trace is the sum of diagonal elements. The exceptional Jordan algebra is $27$ dimensional, just
like the principal fundamental representation of $\mathfrak{e}_6$.\\
In our previous paper \cite{f4} we determined the algebra of derivations $\mathfrak{D}$ and used it to obtain
an irreducible representation of the exceptional algebra $\mathfrak{f}_4$. To obtain a representation of
$\mathfrak{e}_6$ we only need to determine the matrix representation of $R_Y$. This is a very simple task
and we solved it in $\bR^{27}$ with the product inherited from $J_3$ by means of the linear isomorphism
\eqn
&& \Phi : J_3 \longrightarrow \bR^{27} \ , \quad A\mapsto \Phi(A) \ , \cr
&& \Phi(A):=\lp
\begin{array}{c}
a_1 \\
\rho (o_1) \\
\rho (o_2) \\
a_2 \\
\rho (o_3) \\
a_3
\end{array}
\rp \ ,
\feqn
where $A$ is the Jordan matrix
\eqn
\label{A} A=\lp
\begin{array}{ccc}
a_1 & o_1 & o_2 \\
o_1^* & a_2 & o_3 \\
o_2^* & o_3^* & a_3
\end{array}
\rp \ ,
\feqn
and $\rho$ is the linear isomorphism between the octonions $\bO$ and $\bR^8$
given by\footnote{See \cite{f4} for more details.}
\eqn
&& \rho: \bO \longrightarrow \bR \ , \quad o=o^0 +\sum_{i=1}^7 o^i i_i \mapsto \rho (o) \ ,\cr
&& \rho(o):=\lp
\begin{array}{c}
o^0 \\ o^1 \\ o^2 \\ o^3 \\ o^4 \\ o^5 \\ o^6 \\ o^7
\end{array}
\rp \ .
\feqn
After choosing a $26$ dimensional base of traceless Jordan matrices we used the
Mathematica program in App.\ref{app:matrici} to find the matrices which complete the base for $\mathfrak{f}_4$ to
a base for the whole $\mathfrak{e}_6$ algebra. In particular if we work in the real case, we obtain the split form of
$\mathfrak{e}_6$, with signature $(52,26)$. However, multiplying the $26$ added generators by $i$ we find that the algebra remains
real, but the Killing form becomes the compact one.

\section{The construction of the group \boldmath{$E_6$}} \label{group}

\subsection{The generalized Euler parametrization for \boldmath{$E_6$}}
To give an Euler parametrization for $E_6$ we start by choosing its maximal subgroup. It is $H=F_4$, the group generated by the first
$52$ matrices.
Let $\cP$  be the linear complement of $\mathfrak{f}_4$ in $\mathfrak{e}_6$. We then search for a minimal linear subset $V$ of $\cP$, which,
under action of $Ad(F_4)$, generates the whole $\cP$. Looking at the structure constants we see that $V$ can be chosen as the linear space
generated by $c_{53},c_{70}$. Note that they commute.\\
If we then write the general element $g$ of $E_6$ in the form
\eqn
g= \exp (\tilde h) \exp (v) \exp (h)\ , \quad h,\tilde h \in \mathfrak{f}_4\ ,\ v\in V \ ,
\feqn
we have a redundancy of dimension $28$. As argued in \cite{f4} we expect to find a $28$-dimensional subgroup of $F_4$ ,which, acting by
adjunction, defines an automorphism of $V$. This can be done by noticing that $V$ commutes with the first $28$
matrices $c_i$, $i=1,\ldots,28$, which generate an $SO(8)$ subgroup of $F_4$.\\
We found convenient to introduce the change of base
\eqn
&& \tilde c_{53}:= \frac 12 c_{53} +\frac {\sqrt 3}2 c_{70}\ ,\\
&& \tilde c_{70}:=-\frac {\sqrt 3}2 c_{53} +\frac 12 c_{70}\ .
\feqn
Thus we have
$$
g[x_1,\ldots,x_{78}]=B[x_1,\ldots,x_{24}] e^{x_{25}\tilde c_{53}} e^{x_{26}\tilde c_{70}} F_4 [x_{27},\ldots,x_{78}] \ ,
$$
where $B=F_4 / SO(8)$. We chose for $F_4$ the Euler parametrization given in \cite{f4} so that we found for $B$
\eqn
B[x_1,\ldots,x_{24}]=B_{F_4}[x_1,\ldots,x_{16}] B_9[x_{17},\ldots,x_{23}]e^{x_{24}c_{45}} \ ,
\feqn
where
\eqn
&& B_{F_4}[x_1,\ldots,x_{15}]=B_9 [x_1,\ldots,x_7] e^{x_8 c_{45}} B_8 [x_9,\ldots,x_{14}] e^{x_{15} \tilde c_{30}}e^{x_{16} c_{22}} \ ,\\
&& B_9[x_1,\ldots,x_7] =e^{x_1 \tilde c_3} e^{x_2 \tilde c_{16}} e^{x_3 \tilde c_{15}} e^{x_4 \tilde c_{35}} e^{x_5 \tilde c_5}
e^{x_6 \tilde c_1} e^{x_7 \tilde c_{30}}\ ,\\
&& B_8[x_1,\ldots,x_6] =e^{x_1 \tilde c_3} e^{x_2 \tilde c_{16}} e^{x_3 \tilde c_{15}} e^{x_4 \tilde c_{35}} e^{x_5 \tilde c_5} e^{x_6 \tilde c_1} \ ,
\feqn
and the tilded matrices are the ones introduced in \cite{f4}.

\subsection{Determination of the range for the parameters.}
To determine the range of the parameters we will use the topological method
we have developed in \cite{noi}. Let us first determine the volume of $E_6$
by means of the Macdonald formula.

\subsubsection{The volume of \boldmath{$E_6$}.}
The rational homology of the exceptional Lie group $E_6$ is that of a product of odd
dimensional spheres~\cite{hopf},
$H_*(E_6)=H_*(\prod_{i=1}^6 S^{d_i})$, with (\cite{chevalley})
\eqn
d_1=3, \ d_2=9, \ d_3=11, \ d_4=15, \ d_5=17, \ d_6=23\ .
\feqn
The simple roots of $E_6$ are
\eqn
&& r_1 =L_1+L_2 \ ,\\
&& r_2 =L_2-L_1 \ , \\
&& r_3 =L_3-L_2 \ , \\
&& r_4 =L_4-L_3 \ , \\
&& r_5 =L_5-L_4 \ , \\
&& r_6 =\frac {L_1-L_2-L_3-L_4-L_5+\sqrt 3 L_6}2 \ ,
\feqn
where $L_i$, $i=1,\ldots,6$ is an orthonormal base for the Cartan algebra. The volume of the
fundamental region is then
\eqn
Vol(f_{E_6})= \frac 2L \ .
\feqn
Indeed we computed the $36$ positive roots, all having length $\sqrt 2$. They have exactly the structure given in
\cite{fulton-harris}, with $L_i=e_i$, the canonical base of $\mathbb{R}^6$.
The Macdonald formula \cite{mac}, \cite{burgy} gives for the volume of the compact form of $E_6$
\eqn
Vol(E_6) =\frac {\sqrt 3 \cdot 2^{17} \cdot \pi^{42}}{3^{10} \cdot 5^5 \cdot 7^3 \cdot 11} \ .\label{volumeE6}
\feqn

\subsubsection{The invariant measure on \boldmath{$E_6$}.}
With the chosen generalized Euler parametrization, the invariant measure on $E_6$
decomposes into the product of the measure of $F_4$ and the one on $M=E_6/F_4$.
The invariant measure on $F_4$ was computed in \cite{f4} so that we need to compute here
only the induced measure on $M$.\\
Using the notation of \cite{f4}, let us define
\eqn
J_M:=\pi_{\cP} (e^{-x_{26}\tilde c_{70}}e^{-x_{25}\tilde c_{53}} B[x_1,\ldots,x_{24}]^{-1}
d(B[x_1,\ldots,x_{24}] e^{x_{25}\tilde c_{53}}e^{x_{26}\tilde c_{70}})) \ ,
\feqn
where $\pi_\cP$ is the projection on the subspace generated by $c_{j}$, $j=53,\ldots,78$.
The metric induced on $M$ by the bi-invariant metric on $E_6$ is then
\eqn
ds^2_M=-\frac 16 Trace(J_M \otimes J_M) \ ,
\feqn
and the invariant measure on $E_6$ is then
\eqn
d\mu_{E_6}=|\det(J_{Mi}^j)| d\mu_{F_4} \prod_{l=1}^{26} dx_l \ ,
\feqn
where $J_{Mi}^j$ is the $26\times 26$ matrix defined by
\eqn
J_M =\sum_{i,j=1}^{26} J_{Mi}^j c_{i_j} dx^i \ , \feqn
with $c_{i_j}$ a base $\{\tilde c_{53},\tilde c_{70}, c_{54},\ldots,c_{69},c_{71},\ldots, c_{78}\}$ of $\cP$.\\
In order to compute $|det(J_{Mi}^j)|$ it is convenient to introduce the notations
\eqn
&& \omega[x,y,z]=e^{x c_{45}}e^{y \tilde c_{53}}e^{z \tilde c_{70}}\ ,\\
&& J_\omega :=\omega^{-1} d\omega \ ,\\
&& J_9:=B_9^{-1} dB_9 \ ,\\
&& J_{F_4}:= B_{F_4}^{-1} dB_{F_4}\ ,
\feqn
so that
\eqn
&& J_M [x_1,\ldots, x_{26}]\cr
&& \qquad\qquad =\omega[x_{24}, x_{25}, x_{26}]^{-1}  B_9[x_{17},\ldots, x_{23}]^{-1}
J_{F_4}[x_1,\ldots, x_{16}] B_9[x_{17},\ldots, x_{23}]  \omega[x_{24}, x_{25}, x_{26}]\cr
&& \qquad \qquad + \omega[x_{24}, x_{25}, x_{26}]^{-1} J_{9}[x_{17},\ldots, x_{23}] \omega[x_{24}, x_{25}, x_{26}]
+J_\omega[x_{24}, x_{25}, x_{26}]  \ .
\feqn
Some remarks are in order now
\begin{enumerate}
\item the following relations are true
\eqn
e^{-\al \tilde c_{53}} c_L e^{\al \tilde c_{53}} =\cos {\al} \ c_L +\sin \al \ c_{L+26} \ ,
\feqn
if $L=45, \ldots, 52$. Moreover $\tilde c_{70}$ commutes with $c_I$, $I=45,\ldots,52,71,\ldots, 78$ and with $\tilde c_{53}$;
\item $\tilde c_{53}$ and $\tilde c_{70}$ commute with the $so(8)$ algebra generated by the matrices $c_I$, $I=1, \ldots, 21,$ $30, \ldots, 36$;
\item the adjoint action of $e^{x_{24} c_{45}}$ on the above $so(8)$ algebra generates in addition the matrices $c_J$, $J=46, \ldots, 52$;
\item from $J_{F_4} \in {\mathfrak f}_4$ and $B_9 \subset F_4$, it follows that
$$
\tilde J_{F_4} := B_9^{-1} J_{F_4} B_9 \in {\mathfrak f}_4\ ;
$$
\item the adjoint action of $\omega$ on the ${\mathfrak f}_4$ matrices generates all the remaining matrices of ${\mathfrak e}_6$.
In particular, the projection of $\omega^{-1} c_L \omega$, $L=1,\ldots, 52$ on $c_J$, $J=54, \ldots, 69$, is different from zero
only if $L=22, \ldots, 29, 37, \ldots, 44$;
\item the adjoint action of the $SO(8)$ group, corresponding to the above $so(8)$ algebra, gives a rotation both on the indexes
$I=\{ 22, \ldots, 29 \}$ and $J= \{ 37, \ldots, 44 \}$. More precisely
\eqn
&& SO(8)^{-1} c_I SO(8) = {R_I}^L c_L \ , \\
&& SO(8)^{-1} c_J SO(8) = {\tilde R_J}^K c_K \ ,
\feqn
where $L$ runs from $22$ to $29$ and $K$ runs from $37$ to $44$, and $R, \tilde R$ are both orthogonal matrices.
To verify these it suffices to note that
\eqn
&& e^{-x c_A} c_I e^{x c_A} =\cos \frac {x}{2} \ c_I \pm \sin \frac {x}{2} \ c_{I_A} \ , \\
&& e^{-x c_A} c_J e^{x c_A} =\cos \frac {x}{2} \ c_J \pm \sin \frac {x}{2} \ c_{J_A} \ ,
\feqn
where $A=1, \ldots, 21, 30, \ldots, 36$, $I,I_A \in \{ 22, \ldots, 29 \}$, $J,J_A \in \{ 37, \ldots, 44 \}$.
In particular $\det (R\otimes \tilde R) =1$.
\end{enumerate}
{F}rom the first three points, we see that the matrix $J_M$ takes the form
\eqn
M=\left(
\begin{array}{ccc}
A & 0 & 0 \\
* & C & 0 \\
* & * & D
\end{array}
\right) \ ,
\feqn
where on the rows we indicate the coefficients of $dx_I$
with $I=1, \ldots, 26$ starting from the bottom, and on the columns
the projections on $\tilde c_A$, $c_L$, following the order:
$A=53, 70$, $L=71,\ldots, 78, 54, \ldots, 69$. The asterisks indicate
the elements which do not contribute to the determinant:
$$\det J_M= det A \det C \det D\ .$$
In particular, $A$ is a $3 \times 3$ block obtained by projecting $J_\omega$ on $\tilde c_{53},\tilde c_{70}, c_{71}$.
Point~$1$ implies
\eqn
A=\left(
\begin{array}{ccc}
1 & 0 & 0 \\
0 & 1 & 0 \\
0 & 0 & \sin {x_{25}}
\end{array}
\right) \ ,
\feqn
so that
\eqn
\det A=\sin {x_{25}} \ .
\feqn
The $C$ block is a $7 \times 7$ matrix obtained by projecting
$\omega^{-1} J_{9} \; \omega$ on $c_K$, $K=72, \ldots, 78$. From point~$1$ it follows
that it can be obtained by multiplying
the projection of $e^{-x_{24} c_{45}} J_{9} e^{x_{24} c_{45}}$ on $c_L$, $L=46, \ldots, 52$ by $\sin {x_{25}}$. If we call
$\tilde C$ the matrix corresponding to such a projection, we find
$\det C =\sin^7 {x_{25}} \det \tilde C$.
The determinant of $\tilde C$ can be computed directly and gives
\eqn
\det C =\sin x_{20} \cos x_{21} \cos x_{22} \sin^2 x_{22} \sin^2 x_{23} \cos^4 x_{23} \sin^7 x_{24} \sin^7 x_{25} \ .
\feqn
The $D$ block requires some further discussion. It is a $16 \times 16$ matrix obtained
by projecting $\omega^{-1} \tilde J_{B_{F_4}} \omega$
on $c_I$, $I=54, \ldots, 69$. First, from points~$4,5$ of our remarks we see that only the $c_L$ with $L=22, \ldots, 29, 37, \ldots, 44$,
in $\tilde J_{F_4}$ contribute to the determinant. Let us define the $16 \times 16$ matrix $U$ with
\eqn
&& {U_A}^B :=- \frac{1}{6} \tr (\omega^{-1} c_{A} \omega c_B) \ , \\
&& \ \quad A=22, \ldots, 29, 37, \ldots, 44 \ , \\
&& \ \quad B=54, \ldots, 69 \ .
\feqn
Moreover let  $\tilde D$ be  the matrix obtained by projecting $J_{F_4}$ on $c_L$
with $L=22, \ldots, 29, 37, \ldots, 44$, and
\eqn
Q:=\left(
\begin{array}{cc}
R & 0 \\
0 & \tilde R
\end{array}
\right) \ ,
\feqn
where $R$ and $\tilde R$ are the rotation matrices defined at point $6$. Then we can
deduce from points~$4,5$ and $6$ that
\eqn
\det D =\det (U \cdot Q \cdot \tilde D) =\det U \det \tilde D \ .
\feqn
The matrix $U$ and its determinant are easily computed. Indeed, we have
\eqn
&& \omega^{-1} c_{22} \omega =
\cos \left( \frac {x_{24}}{2} \right) \sin \left( \frac {\sqrt 3}2 x_{26} +\frac {x_{25}}2 \right) c_{61} +
\sin \left( \frac {x_{24}}{2} \right) \sin \left( \frac {\sqrt 3}2 x_{26} -\frac {x_{25}}2 \right) c_{69} +\ldots \cr
&& \omega^{-1} c_{23} \omega =
-\cos \left( \frac {x_{24}}{2} \right) \sin \left( \frac {\sqrt 3}2 x_{26} +\frac {x_{25}}2 \right) c_{57} -
\sin \left( \frac {x_{24}}{2} \right) \sin \left( \frac {\sqrt 3}2 x_{26} -\frac {x_{25}}2 \right) c_{65} +\ldots \cr
&& \omega^{-1} c_{24} \omega =
-\cos \left( \frac {x_{24}}{2} \right) \sin \left( \frac {\sqrt 3}2 x_{26} +\frac {x_{25}}2 \right) c_{60} -
\sin \left( \frac {x_{24}}{2} \right) \sin \left( \frac {\sqrt 3}2 x_{26} -\frac {x_{25}}2 \right) c_{68} +\ldots \cr
&& \omega^{-1} c_{25} \omega =
\cos \left( \frac {x_{24}}{2} \right) \sin \left( \frac {\sqrt 3}2 x_{26} +\frac {x_{25}}2 \right) c_{55} +
\sin \left( \frac {x_{24}}{2} \right) \sin \left( \frac {\sqrt 3}2 x_{26} -\frac {x_{25}}2 \right) c_{63} +\ldots \cr
&& \omega^{-1} c_{26} \omega =
-\cos \left( \frac {x_{24}}{2} \right) \sin \left( \frac {\sqrt 3}2 x_{26} +\frac {x_{25}}2 \right) c_{59} -
\sin \left( \frac {x_{24}}{2} \right) \sin \left( \frac {\sqrt 3}2 x_{26} -\frac {x_{25}}2 \right) c_{67} +\ldots \cr
&& \omega^{-1} c_{27} \omega =
\cos \left( \frac {x_{24}}{2} \right) \sin \left( \frac {\sqrt 3}2 x_{26} +\frac {x_{25}}2 \right) c_{58} +
\sin \left( \frac {x_{24}}{2} \right) \sin \left( \frac {\sqrt 3}2 x_{26} -\frac {x_{25}}2 \right) c_{66} +\ldots \cr
&& \omega^{-1} c_{28} \omega =
\cos \left( \frac {x_{24}}{2} \right) \sin \left( \frac {\sqrt 3}2 x_{26} +\frac {x_{25}}2 \right) c_{56} +
\sin \left( \frac {x_{24}}{2} \right) \sin \left( \frac {\sqrt 3}2 x_{26} -\frac {x_{25}}2 \right) c_{64} +\ldots \cr
&& \omega^{-1} c_{29} \omega =
-\cos \left( \frac {x_{24}}{2} \right) \sin \left( \frac {\sqrt 3}2 x_{26} +\frac {x_{25}}2 \right) c_{54} -
\sin \left( \frac {x_{24}}{2} \right) \sin \left( \frac {\sqrt 3}2 x_{26} -\frac {x_{25}}2 \right) c_{62} +\ldots \cr
&& \omega^{-1} c_{37} \omega =
-\sin \left( \frac {x_{24}}{2} \right) \sin \left( \frac {\sqrt 3}2 x_{26} +\frac {x_{25}}2 \right) c_{54} +
\cos \left( \frac {x_{24}}{2} \right) \sin \left( \frac {\sqrt 3}2 x_{26} -\frac {x_{25}}2 \right) c_{62} +\ldots \cr
&& \omega^{-1} c_{38} \omega =
-\sin \left( \frac {x_{24}}{2} \right) \sin \left( \frac {\sqrt 3}2 x_{26} +\frac {x_{25}}2 \right) c_{55} +
\cos \left( \frac {x_{24}}{2} \right) \sin \left( \frac {\sqrt 3}2 x_{26} -\frac {x_{25}}2 \right) c_{63} +\ldots \cr
&& \omega^{-1} c_{39} \omega =
-\sin \left( \frac {x_{24}}{2} \right) \sin \left( \frac {\sqrt 3}2 x_{26} +\frac {x_{25}}2 \right) c_{56} +
\cos \left( \frac {x_{24}}{2} \right) \sin \left( \frac {\sqrt 3}2 x_{26} -\frac {x_{25}}2 \right) c_{64} +\ldots \cr
&& \omega^{-1} c_{40} \omega =
-\sin \left( \frac {x_{24}}{2} \right) \sin \left( \frac {\sqrt 3}2 x_{26} +\frac {x_{25}}2 \right) c_{57} +
\cos \left( \frac {x_{24}}{2} \right) \sin \left( \frac {\sqrt 3}2 x_{26} -\frac {x_{25}}2 \right) c_{65} +\ldots \cr
&& \omega^{-1} c_{41} \omega =
-\sin \left( \frac {x_{24}}{2} \right) \sin \left( \frac {\sqrt 3}2 x_{26} +\frac {x_{25}}2 \right) c_{58} +
\cos \left( \frac {x_{24}}{2} \right) \sin \left( \frac {\sqrt 3}2 x_{26} -\frac {x_{25}}2 \right) c_{66} +\ldots \cr
&& \omega^{-1} c_{42} \omega =
-\sin \left( \frac {x_{24}}{2} \right) \sin \left( \frac {\sqrt 3}2 x_{26} +\frac {x_{25}}2 \right) c_{59} +
\cos \left( \frac {x_{24}}{2} \right) \sin \left( \frac {\sqrt 3}2 x_{26} -\frac {x_{25}}2 \right) c_{67} +\ldots \cr
&& \omega^{-1} c_{43} \omega =
-\sin \left( \frac {x_{24}}{2} \right) \sin \left( \frac {\sqrt 3}2 x_{26} +\frac {x_{25}}2 \right) c_{60} +
\cos \left( \frac {x_{24}}{2} \right) \sin \left( \frac {\sqrt 3}2 x_{26} -\frac {x_{25}}2 \right) c_{68} +\ldots \cr
&& \omega^{-1} c_{44} \omega =
-\sin \left( \frac {x_{24}}{2} \right) \sin \left( \frac {\sqrt 3}2 x_{26} +\frac {x_{25}}2 \right) c_{61} +
\cos \left( \frac {x_{24}}{2} \right) \sin \left( \frac {\sqrt 3}2 x_{26} -\frac {x_{25}}2 \right) c_{69} +\ldots \nonumber
\feqn
where ellipses stay for terms which vanish when projected on $c_B$, $B=54, \ldots, 61, 70, \ldots, 78$.
{From} these it follows
\eqn
\det U=\sin^8 \left( \frac {\sqrt 3}2 x_{26} +\frac {x_{25}}2 \right) \sin^8 \left( \frac {\sqrt 3}2 x_{26} -\frac {x_{25}}2 \right)
\ .
\feqn
At this point we need to compute $\det \tilde D$.
This can be done by noticing that $J_{F_4}$ coincides exactly
with the current $J_{M}$ of $F_4$ in
\cite{f4}.
We then find
\eqn
&& \det \tilde D =2^7 \sin^{15} \frac {x_{16}}2 \cos^7 \frac {x_{16}}2 \sin x_4 \cos x_5 \cos x_6
\sin^2 x_6 \cos^4 x_7 \sin^2 x_7 \sin^7 x_8 \cdot \cr
&& \qquad\ \cdot\sin x_{12} \cos x_{13} \cos x_{14} \sin^2 x_{14} \cos^2 x_{15} \sin^4 x_{15}\ .
\feqn
We can finally write down the invariant measure on the base space:
\eqn
&& d\mu_{M}=2^7 \sin x_4 \cos x_5 \cos x_6 \sin^2 x_6 \cos^4 x_7 \sin^2 x_7 \sin^7 x_8 \cdot \cr
&& \qquad\ \cdot\sin x_{12} \cos x_{13} \cos x_{14} \sin^2 x_{14} \cos^2 x_{15} \sin^4 x_{15}\ \sin^{15} \frac {x_{16}}2
\cos^7 \frac {x_{16}}2  \cdot \cr
&& \qquad\ \cdot \sin x_{20} \cos x_{21} \cos x_{22} \sin^2 x_{22} \sin^2 x_{23} \cos^4 x_{23} \sin^7 x_{24} \cdot \cr
&& \qquad\ \cdot \sin^8 x_{25}
\sin^8 \left( \frac {\sqrt 3}2 x_{26} +\frac {x_{25}}2 \right) \sin^8 \left( \frac {\sqrt 3}2 x_{26} -\frac {x_{25}}2 \right)    \ .
\feqn
Note that the periods of the variables are $4\pi$ so that one should take the range $x_i=[0,4\pi]$ for $i=1,2,3,,9,10,11,17,18,19$.
However, as in \cite{f4}, it is easy to show directly from the parametrization that they can all be restricted
to $[0,2\pi]$. The range of
$x_i$ is then
\eqn
&& x_1 \in [0,2\pi] \ , \quad x_2 \in [0,2\pi] \ , \quad x_3 \in [0,2\pi] \ , \quad x_4 \in [0,\pi] \ , \cr
&& x_5 \in [-\frac \pi2,\frac \pi2] \ , \quad x_6\in [0,\frac \pi2] \ , \quad x_7\in [0,\frac \pi2] \ , \quad x_8\in [0,\pi] \ , \cr
&& x_9 \in [0,2\pi] \ , \quad x_{10} \in [0,2\pi] \ , \quad x_{11} \in [0,2\pi] \ , \quad x_{12} \in [0,\pi] \ , \cr
&& x_{13} \in [-\frac \pi2,\frac \pi2] \ , \quad x_{14}\in [0,\frac \pi2] \ , \quad x_{15}\in [0,\frac \pi2] \ ,
\quad x_{16}\in [0,\pi] \ , \cr
&& x_{17} \in [0,2\pi] \ , \quad x_{18} \in [0,2\pi] \ , \quad x_{19} \in [0,2\pi] \ , \quad x_{20} \in [0,\pi] \ , \cr
&& x_{21} \in [-\frac \pi2,\frac \pi2] \ , \quad x_{22}\in [0,\frac \pi2] \ , \quad x_{23}\in [0,\frac \pi2] \ , \quad x_{24}\in [0,\pi] \ , \cr
&& x_{25}\in [0,\frac \pi2] \ , \quad -\frac {x_{25}}{\sqrt 3} \leq x_{26} \leq \frac {x_{25}}{\sqrt 3} \ .
\feqn
The remaining parameters $x_j$, $j=27,\ldots,78$, will run over the range for $F_4$, as given in \cite{f4}.
The volume of the whole closed cycle $V$ so obtained is then
\eqn
Vol (V)=Vol(F_4)\int_R d\mu_M =\frac {\sqrt 3 \cdot 2^{17} \cdot \pi^{42}}{3^{10} \cdot 5^5 \cdot 7^3 \cdot 11} \ ,
\feqn
where $R$ is the range of parameters $x_i$, $i=1,\ldots,26$.
This is the volume of $E_6$, so that we cover the group exactly once.\footnote{Obviously there is a
subset of vanishing measure which is multiply covered.}

\section{Conclusions.}
In this paper we have performed an explicit construction for the compact form of the
simple Lie group $E_6$. This is particularly interesting, because recently it has
been argued that this group could be the most promising for unification in
GUT theories~\cite{caramor,das}.
To parameterize the group we have used the generalized Euler angles method, a technique
we introduced in \cite{noi,f4} to give the most simple expression for the invariant
measure on the group, while at the same time still being able to provide an explicit
expression for the range of the parameters. Both these requirements are necessary
in order to minimize the computation power needed for computer simulations,
for example, of lattice models. \\
Our results can be easily extended to the GUT group $E_6^4 \rtimes S_4$. Also, a modified
Euler parametrization could be applied to evidentiate the subgroups related
to the correct symmetry breaking, see \cite{caramor}. Finally, our parametrization
could be used for a straightforward geometrical analysis of the exceptional Lie
group $E_6$ and its quotients.


\subsection*{Acknowledgments}
SC is grateful to Giuseppe Berrino and Stefano Pigola for helpful comments.
BLC would like to thank O. Ganor for useful discussions.
This work has been supported in part by the Spanish Ministry of Education and Science (grant AP2005-5201) and in part
by the Director, Office of Science, Office of High Energy and Nuclear Physics of the U.S. Department of Energy
under Contract DE-AC02-05CH11231.

\newpage
\begin{appendix}
\section{The \boldmath{$\mathfrak{e}_6$} matrices.} \label{app:matrici}
The matrices we found using Mathematica, and orthonormalized with
respect to the scalar product $\langle a , b\rangle :=-\frac 16
Trace (ab)$, was computed by means of the followings programs.

\subsection*{The Mathematica program.}
\begin{eqnarray*}
\rm
&& \%\%\%\% \mbox{    the octonionic products}\\
&& \QQ[1,1]=e[1]; \qquad \QQ[1,2]=e[2]; \qquad \QQ[1,3]=e[3];\\
&& \QQ[1,4]=e[4]; \qquad \QQ[1,5]=e[5]; \qquad \QQ[1,6]=e[6];\\
&& \QQ[1,7]=e[7]; \qquad \QQ[1,8]=e[8]; \qquad \QQ[2,1]=e[2];\\
&& \QQ[2,2]=-e[1]; \qquad \QQ[2,3]=e[5]; \qquad \QQ[2,4]=e[8];\\
&& \QQ[2,5]=-e[3]; \qquad \QQ[2,6]=e[7]; \qquad \QQ[2,7]=-e[6];\\
&& \QQ[2,8]=-e[4]; \qquad \QQ[3,1]=e[3]; \qquad \QQ[3,2]=-e[5];\\
&& \QQ[3,3]=-e[1]; \qquad \QQ[3,4]=e[6]; \qquad \QQ[3,5]=e[2];\\
&& \QQ[3,6]=-e[4]; \qquad \QQ[3,7]=e[8]; \qquad \QQ[3,8]=-e[7];\\
&& \QQ[4,1]=e[4]; \qquad \QQ[4,2]=-e[8]; \qquad \QQ[4,3]=-e[6];\\
&& \QQ[4,4]=-e[1]; \qquad \QQ[4,5]=e[7]; \qquad \QQ[4,6]=e[3];\\
&& \QQ[4,7]=-e[5]; \qquad \QQ[4,8]=e[2]; \qquad \QQ[5,1]=e[5];\\
&& \QQ[5,2]=e[3]; \qquad \QQ[5,3]=-e[2]; \qquad \QQ[5,4]=-e[7];\\
&& \QQ[5,5]=-e[1]; \qquad \QQ[5,6]=e[8]; \qquad \QQ[5,7]=e[4];\\
&& \QQ[5,8]=-e[6]; \qquad \QQ[6,1]=e[6]; \qquad \QQ[6,2]=-e[7];\\
&& \QQ[6,3]=e[4]; \qquad \QQ[6,4]=-e[3]; \qquad \QQ[6,5]=-e[8];\\
&& \QQ[6,6]=-e[1]; \qquad \QQ[6,7]=e[2]; \qquad \QQ[6,8]=e[5];\\
&& \QQ[7,1]=e[7]; \qquad \QQ[7,2]=e[6]; \qquad \QQ[7,3]=-e[8];\\
&& \QQ[7,4]=e[5]; \qquad \QQ[7,5]=-e[4]; \qquad \QQ[7,6]=-e[2];\\
&& \QQ[7,7]=-e[1]; \qquad \QQ[7,8]=e[3]; \qquad \QQ[8,1]=e[8];\\
&& \QQ[8,2]=e[4]; \qquad \QQ[8,3]=e[7]; \qquad \QQ[8,4]=-e[2];\\
&& \QQ[8,5]=e[6]; \qquad \QQ[8,6]=-e[5]; \qquad \QQ[8,7]=-e[3];\\
&& \QQ[8,8]=-e[1];\\
&& \%\%\%\% \mbox{    the Jordan algebra product}\\
&& \Qm[x_\_,y_\_] :=\SSu[\SSu[x[[i]]y[[j]]\QQ[i,j],\{i,8\}],\{j,8\}]\\
&& \QP[x_\_,y_\_] :=\left\{\Coe[\Qm[x,y],e[1]],\ \Coe[\Qm[x,y],e[2]], \right.\\
&& \qquad \Coe [\Qm[x,y],e[3]],\ \Coe [\Qm[x,y],e[4]],\ \Coe [\Qm[x,y],e[5]], \\
&& \left. \qquad \Coe [\Qm[x,y],e[6]],\ \Coe [\Qm[x,y],e[7]],\ \Coe [\Qm[x,y],e[8]]\right\}\\
\end{eqnarray*}
\begin{eqnarray*}
&& \Octp[a_\_,b_\_]:= \left\{\left\{\SSu [\QP[\Part[\Part[a,1],i],\Part[\Part[b,i],1]],\{i,3\}]\right.\right. ,\\
&& \qquad \SSu [\QP[\Part[\Part[a,1],i],\Part[\Part[b,i],2]],\{i,3\}],\\
&& \left. \qquad \SSu [\QP[\Part[\Part[a,1],i],\Part[\Part[b,i],3]],\{i,3\}]\right\} ,\\
&& \quad\ \ \left\{ \SSu [\QP[\Part[\Part[a,2],i],\Part[\Part[b,i],1]],\{i,3\}], \right. \\
&& \qquad \SSu [\QP[\Part[\Part[a,2],i],\Part[\Part[b,i],2]],\{i,3\}],\\
&& \left. \qquad \SSu [\QP[\Part[\Part[a,2],i],\Part[\Part[b,i],3]],\{i,3\}]\right\} ,\\
&& \quad\ \ \left\{ \SSu [\QP[\Part[\Part[a,3],i],\Part[\Part[b,i],1]],\{i,3\}], \right. \\
&& \qquad \SSu [\QP[\Part[\Part[a,3],i],\Part[\Part[b,i],2]],\{i,3\}],\\
&& \left. \left. \qquad \SSu [\QP[\Part[\Part[a,3],i],\Part[\Part[b,i],3]],\{i,3\}]\right\}\right\}\\
&& \Octps[a_\_,b_\_]:=1/2 (\Octp[a,b]+\Octp[b,a])\\
\end{eqnarray*}
\begin{eqnarray*}
&& \%\%\%\% \mbox{    correspondence between the Jordan algebra and $\bR^{27}$}\\
&& {\rm FF}[{\rm AA}_\_]:=\left\{ \Part[\{\Part[\Part[\Part[{\rm AA},1],\ 1],\ 1]\},\ 1],\right.\\
&& \qquad \Part[\{\Part[\Part[\Part[{\rm AA},1],\ 2],\ 1]\},\ 1],\ \Part[\{\Part[\Part[\Part[{\rm AA},1],\ 2],\ 2]\},\ 1],\\
&& \qquad \Part[\{\Part[\Part[\Part[{\rm AA},1],\ 2],\ 3]\},\ 1],\ \Part[\{\Part[\Part[\Part[{\rm AA},1],\ 2],\ 4]\},\ 1],\\
&& \qquad \Part[\{\Part[\Part[\Part[{\rm AA},1],\ 2],\ 5]\},\ 1],\ \Part[\{\Part[\Part[\Part[{\rm AA},1],\ 2],\ 6]\},\ 1],\\
&& \qquad \Part[\{\Part[\Part[\Part[{\rm AA},1],\ 2],\ 7]\},\ 1],\ \Part[\{\Part[\Part[\Part[{\rm AA},1],\ 2],\ 8]\},\ 1],\\
&& \qquad \Part[\{\Part[\Part[\Part[{\rm AA},1],\ 3],\ 1]\},\ 1],\ \Part[\{\Part[\Part[\Part[{\rm AA},1],\ 3],\ 2]\},\ 1],\\
&& \qquad \Part[\{\Part[\Part[\Part[{\rm AA},1],\ 3],\ 3]\},\ 1],\ \Part[\{\Part[\Part[\Part[{\rm AA},1],\ 3],\ 4]\},\ 1],\\
&& \qquad \Part[\{\Part[\Part[\Part[{\rm AA},1],\ 3],\ 5]\},\ 1],\ \Part[\{\Part[\Part[\Part[{\rm AA},1],\ 3],\ 6]\},\ 1],\\
&& \qquad \Part[\{\Part[\Part[\Part[{\rm AA},1],\ 3],\ 7]\},\ 1],\ \Part[\{\Part[\Part[\Part[{\rm AA},1],\ 3],\ 8]\},\ 1],\\
&& \qquad \Part[\{\Part[\Part[\Part[{\rm AA},2],\ 2],\ 1]\},\ 1],\ \Part[\{\Part[\Part[\Part[{\rm AA},2],\ 3],\ 1]\},\ 1],\\
&& \qquad \Part[\{\Part[\Part[\Part[{\rm AA},2],\ 3],\ 2]\},\ 1],\ \Part[\{\Part[\Part[\Part[{\rm AA},2],\ 3],\ 3]\},\ 1],\\
&& \qquad \Part[\{\Part[\Part[\Part[{\rm AA},2],\ 3],\ 4]\},\ 1],\ \Part[\{\Part[\Part[\Part[{\rm AA},2],\ 3],\ 5]\},\ 1],\\
&& \qquad \Part[\{\Part[\Part[\Part[{\rm AA},2],\ 3],\ 6]\},\ 1],\ \Part[\{\Part[\Part[\Part[{\rm AA},2],\ 3],\ 7]\},\ 1],\\
&& \left.\qquad \Part[\{\Part[\Part[\Part[{\rm AA},2],\ 3],\ 8]\},\ 1],\ \Part[\{\Part[\Part[\Part[{\rm AA},3],\ 3],\ 1]\},\ 1]\right\}\\
\end{eqnarray*}
\begin{eqnarray*}
&& {\rm FFi}[{\rm vv}_\_]:=\\
&& \ \ \left\{ \left\{ \{ \Part[{\rm vv},\ 1],\ 0,\ 0,\ 0,\ 0,\ 0,\ 0,\ 0\},\ \left\{ \Part[{\rm vv},\ 2],\ \Part[{\rm vv},\ 3],\
   \Part[{\rm vv},\ 4],\ \Part[{\rm vv},\ 5], \right. \right. \right. \\
&& \left. \qquad  \Part[{\rm vv},\ 6],\ \Part[{\rm vv},\ 7],\ \Part[{\rm vv},\ 8],\ \Part[{\rm vv},\ 9]\right\},\ \left\{
   \Part[{\rm vv},\ 10],\ \Part[{\rm vv},\ 11], \right. \\
&& \left.\left. \qquad  \!\Part[{\rm vv},\ 12],\ \Part[{\rm vv},\ 13],\ \Part[{\rm vv},\ 14],\ \Part[{\rm vv},\ 15],\
   \Part[{\rm vv},\ 16],\ \Part[{\rm vv},\ 17]\right\}\right\},\\
&& \quad \left\{\left\{ \Part[{\rm vv},\ 2],\ -\Part[{\rm vv},\ 3],\ -\Part[{\rm vv},\ 4],\ -\Part[{\rm vv},\ 5],\ -\Part[{\rm vv},\ 6],
   \right.\right. \\
&& \left. \qquad  -\Part[{\rm vv},\ 7],\ -\Part[{\rm vv},\ 8],\ -\Part[{\rm vv},\ 9]\right\},\ \{
   \Part[{\rm vv},\ 18],\  0,\ 0,\ 0,\ 0,\ 0,\ 0,\ 0\}, \\
&& \quad \ \ \left\{ \Part[{\rm vv},\ 19],\ \Part[{\rm vv},\ 20],\ \Part[{\rm vv},\ 21],\ \Part[{\rm vv},\ 22],\right.\\
&& \left.\left. \qquad \! \Part[{\rm vv},\ 23],\ \Part[{\rm vv},\ 24],\ \Part[{\rm vv},\ 25],\ \Part[{\rm vv},\ 26]\right\}\right\},\\
\end{eqnarray*}
\begin{eqnarray*}
&& \quad \left\{\left\{ \Part[{\rm vv},\ 10],\ -\Part[{\rm vv},\ 11],\ -\Part[{\rm vv},\ 12],\ -\Part[{\rm vv},\ 13], \right.\right. \\
&& \left. \qquad  -\Part[{\rm vv},\ 14],\ -\Part[{\rm vv},\ 15],\ -\Part[{\rm vv},\ 16],\ -\Part[{\rm vv},\ 17] \right\},\\
&& \quad \ \ \left\{ \Part[{\rm vv},\ 19],\ -\Part[{\rm vv},\ 20],\ -\Part[{\rm vv},\ 21],\ -\Part[{\rm vv},\ 22],\ -\Part[{\rm vv},\ 23],
   \right.\\
&& \left.\left.\left. \qquad \! -\Part[{\rm vv},\ 24],\ -\Part[{\rm vv},\ 25],\ -\Part[{\rm vv},\ 26]\right\},\ \{
   \Part[{\rm vv},\ 27],\  0,\ 0,\ 0,\ 0,\ 0,\ 0,\ 0\}\right\}\right\}\\
\end{eqnarray*}
\begin{eqnarray*}
&& \%\%\%\% \mbox{     construction of the matrices}\\
&& {\rm MT=\{\ \{\{mt[1],\ 0,\ 0,\ 0,\ 0,\ 0,\ 0,\ 0\},\ \{mt[2],\ mt[3],\ mt[4],\ mt[5],\ mt[6],\ mt[7],\ mt[8], }\\
&& \qquad \quad\ {\rm mt[9]\},\ \{mt[10],\ mt[11],\ mt[12],\ mt[13],\ mt[14],\ mt[15],\ mt[16],\ mt[17]\}\},}\\
&& \qquad\ {\rm \{\{ mt[2],\ -mt[3],\ -mt[4],\ -mt[5],\ -mt[6],\ -mt[7],\ -mt[8],\ -mt[9]\},\ \{mt[18],\ 0,\ 0,}\\
&& \qquad \quad\ {\rm 0,\ 0,\ 0,\ 0,\ 0\},\ \{mt[19],\ mt[20],\ mt[21],\ mt[22],\ mt[23],\ mt[24],\ mt[25],\ mt[26]\}\},}\\
&& \qquad\ {\rm \{\{ mt[10],\ -mt[11],\ -mt[12],\ -mt[13],\ -mt[14],\ -mt[15],\ -mt[16],\ -mt[17]\},}\\
&& \qquad \quad\ {\rm \{mt[19],\ -mt[20],\ -mt[21],\ -mt[22],\ -mt[23],\ -mt[24],\ -mt[25],\ -mt[26]\},}\\
&& \qquad \quad\ {\rm \{-mt[1]-mt[18],\ 0,\ 0,\ 0,\ 0,\ 0,\ 0,\ 0\}\}\};}
\end{eqnarray*}
\begin{eqnarray*}
&& {\rm h=Array[hh,\ 27]};\\
&& {\rm M}=\Array[{\rm mm}, \{27,27\}];\\
&& {\rm imm=FF[OctPS[MT,FFi[h]]]};\\
&& {\rm Do[Do[mm[i,\ j]\ =Coefficient[Part[imm,\ i],\ Part[h,\ j]],\ \{i,27\}],\ \{j,27\}]}\\
&& {\rm Do[econi[j]= D[M,\ mt[j]],\ \{j,\ 26\}];}\\
&& {\rm Mno[1] = econi[1];}\\
&& {\rm Do[Do[}\\
&& \quad {\rm AA[j,\ i] = 0,\ \{i,\ 26\}],\{j,\ 26 \}]}\\
&& {\rm Do[Do[AA[j,\ i] = -Tr[econi[j]\ .Mno[i]]/Tr[Mno[i]\ .Mno[i]],\ \{i,\ j-1\}];}\\
&& \quad {\rm Mno[j] = econi[j] + Sum[AA[j,\ i] Mno[i],\ \{i,\ j-1\}],\ \{j,\ 2,\ 26\}]}\\
&& {\rm Do[Mnno[i] =-\sqrt 6 Mno[i]/Sqrt[Tr[Mno[i]\ .Mno[i]]],\ \{i,\ 26\}];}
\end{eqnarray*}
{\footnotesize
\begin{eqnarray*}
&& \%\%\% \mbox{     rotation to give the irreducible $26$ representation       }\\
&& XXX=\{\{1,\ 0,\ 0,\ 0,\ 0,\ 0,\ 0,\ 0,\ 0,\ 0,\ 0,\ 0,\ 0,\ 0,\ 0,\ 0,\ 0,\ -1,\ 0,\ 0,\ 0,\ 0,\ 0,\ 0,\ 0,\ 0,\ 0 \}/{\rm Sqrt}[2],\\
&& \qquad\qquad\  \{0,\ 1,\ 0,\ 0,\ 0,\ 0,\ 0,\ 0,\ 0,\ 0,\ 0,\ 0,\ 0,\ 0,\ 0,\ 0,\ 0,\ 0,\ 0,\ 0,\ 0,\ 0,\ 0,\ 0,\ 0,\ 0,\ 0\}, \\
&& \qquad\qquad\  \{0,\ 0,\ 1,\ 0,\ 0,\ 0,\ 0,\ 0,\ 0,\ 0,\ 0,\ 0,\ 0,\ 0,\ 0,\ 0,\ 0,\ 0,\ 0,\ 0,\ 0,\ 0,\ 0,\ 0,\ 0,\ 0,\ 0\}, \\
&& \qquad\qquad\  \{0,\ 0,\ 0,\ 1,\ 0,\ 0,\ 0,\ 0,\ 0,\ 0,\ 0,\ 0,\ 0,\ 0,\ 0,\ 0,\ 0,\ 0,\ 0,\ 0,\ 0,\ 0,\ 0,\ 0,\ 0,\ 0,\ 0\}, \\
&& \qquad\qquad\  \{0,\ 0,\ 0,\ 0,\ 1,\ 0,\ 0,\ 0,\ 0,\ 0,\ 0,\ 0,\ 0,\ 0,\ 0,\ 0,\ 0,\ 0,\ 0,\ 0,\ 0,\ 0,\ 0,\ 0,\ 0,\ 0,\ 0\}, \\
&& \qquad\qquad\  \{0,\ 0,\ 0,\ 0,\ 0,\ 1,\ 0,\ 0,\ 0,\ 0,\ 0,\ 0,\ 0,\ 0,\ 0,\ 0,\ 0,\ 0,\ 0,\ 0,\ 0,\ 0,\ 0,\ 0,\ 0,\ 0,\ 0\}, \\
&& \qquad\qquad\  \{0,\ 0,\ 0,\ 0,\ 0,\ 0,\ 1,\ 0,\ 0,\ 0,\ 0,\ 0,\ 0,\ 0,\ 0,\ 0,\ 0,\ 0,\ 0,\ 0,\ 0,\ 0,\ 0,\ 0,\ 0,\ 0,\ 0\},\\
&& \qquad\qquad\  \{0,\ 0,\ 0,\ 0,\ 0,\ 0,\ 0,\ 1,\ 0,\ 0,\ 0,\ 0,\ 0,\ 0,\ 0,\ 0,\ 0,\ 0,\ 0,\ 0,\ 0,\ 0,\ 0,\ 0,\ 0,\ 0,\ 0\}, \\
&& \qquad\qquad\  \{0,\ 0,\ 0,\ 0,\ 0,\ 0,\ 0,\ 0,\ 1,\ 0,\ 0,\ 0,\ 0,\ 0,\ 0,\ 0,\ 0,\ 0,\ 0,\ 0,\ 0,\ 0,\ 0,\ 0,\ 0,\ 0,\ 0\}, \\
&& \qquad\qquad\  \{0,\ 0,\ 0,\ 0,\ 0,\ 0,\ 0,\ 0,\ 0,\ 1,\ 0,\ 0,\ 0,\ 0,\ 0,\ 0,\ 0,\ 0,\ 0,\ 0,\ 0,\ 0,\ 0,\ 0,\ 0,\ 0,\ 0\}, \\
&& \qquad\qquad\  \{0,\ 0,\ 0,\ 0,\ 0,\ 0,\ 0,\ 0,\ 0,\ 0,\ 1,\ 0,\ 0,\ 0,\ 0,\ 0,\ 0,\ 0,\ 0,\ 0,\ 0,\ 0,\ 0,\ 0,\ 0,\ 0,\ 0\},\\
&& \qquad\qquad\  \{0,\ 0,\ 0,\ 0,\ 0,\ 0,\ 0,\ 0,\ 0,\ 0,\ 0,\ 1,\ 0,\ 0,\ 0,\ 0,\ 0,\ 0,\ 0,\ 0,\ 0,\ 0,\ 0,\ 0,\ 0,\ 0,\ 0\}, \\
&& \qquad\qquad\  \{0,\ 0,\ 0,\ 0,\ 0,\ 0,\ 0,\ 0,\ 0,\ 0,\ 0,\ 0,\ 1,\ 0,\ 0,\ 0,\ 0,\ 0,\ 0,\ 0,\ 0,\ 0,\ 0,\ 0,\ 0,\ 0,\ 0\}, \\
&& \qquad\qquad\  \{0,\ 0,\ 0,\ 0,\ 0,\ 0,\ 0,\ 0,\ 0,\ 0,\ 0,\ 0,\ 0,\ 1,\ 0,\ 0,\ 0,\ 0,\ 0,\ 0,\ 0,\ 0,\ 0,\ 0,\ 0,\ 0,\ 0\}, \\
&& \qquad\qquad\  \{0,\ 0,\ 0,\ 0,\ 0,\ 0,\ 0,\ 0,\ 0,\ 0,\ 0,\ 0,\ 0,\ 0,\ 1,\ 0,\ 0,\ 0,\ 0,\ 0,\ 0,\ 0,\ 0,\ 0,\ 0,\ 0,\ 0\}, \\
&& \qquad\qquad\  \{0,\ 0,\ 0,\ 0,\ 0,\ 0,\ 0,\ 0,\ 0,\ 0,\ 0,\ 0,\ 0,\ 0,\ 0,\ 1,\ 0,\ 0,\ 0,\ 0,\ 0,\ 0,\ 0,\ 0,\ 0,\ 0,\ 0\}, \\
&& \qquad\qquad\  \{0,\ 0,\ 0,\ 0,\ 0,\ 0,\ 0,\ 0,\ 0,\ 0,\ 0,\ 0,\ 0,\ 0,\ 0,\ 0,\ 1,\ 0,\ 0,\ 0,\ 0,\ 0,\ 0,\ 0,\ 0,\ 0,\ 0\}, \\
&& \qquad\qquad\  \{1,\ 0,\ 0,\ 0,\ 0,\ 0,\ 0,\ 0,\ 0,\ 0,\ 0,\ 0,\ 0,\ 0,\ 0,\ 0,\ 0,\ 1,\ 0,\ 0,\ 0,\ 0,\ 0,\ 0,\ 0,\ 0,\ -2\}/{\rm Sqrt}[6], \\
&& \qquad\qquad\  \{0,\ 0,\ 0,\ 0,\ 0,\ 0,\ 0,\ 0,\ 0,\ 0,\ 0,\ 0,\ 0,\ 0,\ 0,\ 0,\ 0,\ 0,\ 1,\ 0,\ 0,\ 0,\ 0,\ 0,\ 0,\ 0,\ 0\}, \\
&& \qquad\qquad\  \{0,\ 0,\ 0,\ 0,\ 0,\ 0,\ 0,\ 0,\ 0,\ 0,\ 0,\ 0,\ 0,\ 0,\ 0,\ 0,\ 0,\ 0,\ 0,\ 1,\ 0,\ 0,\ 0,\ 0,\ 0,\ 0,\ 0\}, \\
&& \qquad\qquad\  \{0,\ 0,\ 0,\ 0,\ 0,\ 0,\ 0,\ 0,\ 0,\ 0,\ 0,\ 0,\ 0,\ 0,\ 0,\ 0,\ 0,\ 0,\ 0,\ 0,\ 1,\ 0,\ 0,\ 0,\ 0,\ 0,\ 0\}, \\
&& \qquad\qquad\  \{0,\ 0,\ 0,\ 0,\ 0,\ 0,\ 0,\ 0,\ 0,\ 0,\ 0,\ 0,\ 0,\ 0,\ 0,\ 0,\ 0,\ 0,\ 0,\ 0,\ 0,\ 1,\ 0,\ 0,\ 0,\ 0,\ 0\}, \\
&& \qquad\qquad\  \{0,\ 0,\ 0,\ 0,\ 0,\ 0,\ 0,\ 0,\ 0,\ 0,\ 0,\ 0,\ 0,\ 0,\ 0,\ 0,\ 0,\ 0,\ 0,\ 0,\ 0,\ 0,\ 1,\ 0,\ 0,\ 0,\ 0\}, \\
&& \qquad\qquad\  \{0,\ 0,\ 0,\ 0,\ 0,\ 0,\ 0,\ 0,\ 0,\ 0,\ 0,\ 0,\ 0,\ 0,\ 0,\ 0,\ 0,\ 0,\ 0,\ 0,\ 0,\ 0,\ 0,\ 1,\ 0,\ 0,\ 0\}, \\
&& \qquad\qquad\  \{0,\ 0,\ 0,\ 0,\ 0,\ 0,\ 0,\ 0,\ 0,\ 0,\ 0,\ 0,\ 0,\ 0,\ 0,\ 0,\ 0,\ 0,\ 0,\ 0,\ 0,\ 0,\ 0,\ 0,\ 1,\ 0,\ 0\}, \\
&& \qquad\qquad\  \{0,\ 0,\ 0,\ 0,\ 0,\ 0,\ 0,\ 0,\ 0,\ 0,\ 0,\ 0,\ 0,\ 0,\ 0,\ 0,\ 0,\ 0,\ 0,\ 0,\ 0,\ 0,\ 0,\ 0,\ 0,\ 1,\ 0\}, \\
&& \qquad\qquad\  \{1,\ 0,\ 0,\ 0,\ 0,\ 0,\ 0,\ 0,\ 0,\ 0,\ 0,\ 0,\ 0,\ 0,\ 0,\ 0,\ 0,\ 1,\ 0,\ 0,\ 0,\ 0,\ 0,\ 0,\ 0,\ 0,\ 1\}/{\rm Sqrt}[3] \};
\end{eqnarray*}
}
\begin{eqnarray*}
&& \Do[cc[i+52]=XXX.Mnno[i].{\rm Transpose}[XXX],\ \{i,\ 26\}];
\end{eqnarray*}
\

\noindent The matrices cc[i], ${\rm i}=53,\ldots,78$ so obtained, must be added to the $52$ matrices cc[i] determined in
\cite{f4} to give a base of the $\mathfrak {e}_6$ algebra.

\section{Structure constants.} \label{app:struttura}
The structure constants ${s_{IJ}}^K$ are defined by
$[C_I,C_J]=\sum_{K=1}^{78}{s_{IJ}}^K C_K$. We checked that the
coefficients $s_{IJK}:={s_{IJ}}^K$ are completely antisymmetric in
the indices. We write only the non-vanishing terms, up to symmetries, which are not yet written in \cite{f4}:
\arr
\bea{cccc}
s_{1,54,55}=\dem \ , & s_{1,56,58}=-\dem \ , & s_{1,57,61}=-\dem \ , & s_{1,59,60}=-\dem \ , \\
s_{1,62,63}=-\dem \ , & s_{1,64,66}=\dem \ , & s_{1,65,69}=\dem \ , & s_{1,67,68}=\dem \ , \\
s_{1,71,72}=-1 \ , & s_{2,54,56}=\dem \ , & s_{2,55,58}=\dem \ , & s_{2,57,59}=-\dem \ , \\
s_{2,60,61}=-\dem \ , & s_{2,62,64}=-\dem \ , & s_{2,63,66}=-\dem \ , & s_{2,65,67}=\dem \ , \\
s_{2,68,69}=\dem \ , & s_{2,71,73}=-1 \ , & s_{3,54,58}=\dem \ , & s_{3,55,56}=-\dem \ , \\
s_{3,57,60}=-\dem \ , & s_{3,59,61}=\dem \ , & s_{3,62,66}=\dem \ , & s_{3,63,64}=-\dem \ , \\
s_{3,65,68}=-\dem \ , & s_{3,67,69}=\dem \ , & s_{3,72,73}=-1 \ , & s_{4,54,57}=\dem \ , \\
s_{4,55,61}=\dem \ , & s_{4,56,59}=\dem \ , & s_{4,58,60}=-\dem \ , & s_{4,62,65}=-\dem \ , \\
s_{4,63,69}=-\dem \ , & s_{4,64,67}=-\dem \ , & s_{4,66,68}=\dem \ , & s_{4,71,74}=-1 \ , \\
s_{5,54,61}=\dem \ , & s_{5,55,57}=-\dem \ , & s_{5,56,60}=\dem \ , & s_{5,58,59}=\dem \ , \\
s_{5,62,69}=\dem \ , & s_{5,63,65}=-\dem \ , & s_{5,64,68}=\dem \ , & s_{5,66,67}=\dem \ , \\
s_{5,72,74}=-1 \ , & s_{6,54,59}=\dem \ , & s_{6,55,60}=-\dem \ , & s_{6,56,57}=-\dem \ , \\
s_{6,58,61}=-\dem \ , & s_{6,62,67}=\dem \ , & s_{6,63,68}=-\dem \ , & s_{6,64,65}=-\dem \ , \\
s_{6,66,69}=-\dem \ , & s_{6,73,74}=-1 \ , & s_{7,54,58}=\dem \ , & s_{7,55,56}=-\dem \ , \\
s_{7,57,60}=\dem \ , & s_{7,59,61}=-\dem \ , & s_{7,62,66}=-\dem \ , & s_{7,63,64}=\dem \ , \\
s_{7,65,68}=-\dem \ , & s_{7,67,69}=\dem \ , & s_{7,71,75}=-1 \ , & s_{8,54,56}=-\dem \ , \\
s_{8,55,58}=-\dem \ , & s_{8,57,59}=-\dem \ , & s_{8,60,61}=-\dem \ , & s_{8,62,64}=-\dem \ , \\
s_{8,63,66}=-\dem \ , & s_{8,65,67}=-\dem \ , & s_{8,68,69}=-\dem \ , & s_{8,72,75}=-1 \ , \\
s_{9,54,55}=\dem \ , & s_{9,56,58}=-\dem \ , & s_{9,57,61}=\dem \ , & s_{9,59,60}=\dem \ , \\
s_{9,62,63}=\dem \ , & s_{9,64,66}=-\dem \ , & s_{9,65,69}=\dem \ , & s_{9,67,68}=\dem \ , \\
s_{9,73,75}=-1 \ , & s_{10,54,60}=\dem \ , & s_{10,55,59}=\dem \ , & s_{10,56,61}=-\dem \ , \\
s_{10,57,58}=-\dem \ , & s_{10,62,68}=\dem \ , & s_{10,63,67}=\dem \ , & s_{10,64,69}=-\dem \ , \\
s_{10,65,66}=-\dem \ , & s_{10,74,75}=-1 \ , & s_{11,54,59}=\dem \ , & s_{11,55,60}=\dem \ , \\
s_{11,56,57}=-\dem \ , & s_{11,58,61}=\dem \ , & s_{11,62,67}=-\dem \ , & s_{11,63,68}=-\dem \ , \\
s_{11,64,65}=\dem \ , & s_{11,66,69}=-\dem \ , & s_{11,71,76}=-1 \ , & s_{12,54,60}=\dem \ , \\
s_{12,55,59}=-\dem \ , & s_{12,56,61}=-\dem \ , & s_{12,57,58}=\dem \ , & s_{12,62,68}=\dem \ , \\
s_{12,63,67}=-\dem \ , & s_{12,64,69}=-\dem \ , & s_{12,65,66}=\dem \ , & s_{12,72,76}=-1 \ , \\
s_{13,54,57}=-\dem \ , & s_{13,55,61}=\dem \ , & s_{13,56,59}=-\dem \ , & s_{13,58,60}=-\dem \ , \\
s_{13,62,65}=-\dem \ , & s_{13,63,69}=\dem \ , & s_{13,64,67}=-\dem \ , & s_{13,66,68}=\dem \ , \\
s_{13,73,76}=-1 \ , & s_{14,54,56}=\dem \ , & s_{14,55,58}=-\dem \ , & s_{14,57,59}=-\dem \ , \\
s_{14,60,61}=\dem \ , & s_{14,62,64}=\dem \ , & s_{14,63,66}=-\dem \ , & s_{14,65,67}=-\dem \ , \\
s_{14,68,69}=\dem \ , & s_{14,74,76}=-1 \ , & s_{15,54,61}=\dem \ , & s_{15,55,57}=\dem \ , \\
s_{15,56,60}=\dem \ , & s_{15,58,59}=-\dem \ , & s_{15,62,69}=\dem \ , & s_{15,63,65}=\dem \ , \\
s_{15,64,68}=\dem \ , & s_{15,66,67}=-\dem \ , & s_{15,75,76}=-1 \ , & s_{16,54,60}=\dem \ , \\
s_{16,55,59}=-\dem \ , & s_{16,56,61}=\dem \ , & s_{16,57,58}=-\dem \ , & s_{16,62,68}=-\dem \ , \\
s_{16,63,67}=\dem \ , & s_{16,64,69}=-\dem \ , & s_{16,65,66}=\dem \ , & s_{16,71,77}=-1 \ , \\
s_{17,54,59}=-\dem \ , & s_{17,55,60}=-\dem \ , & s_{17,56,57}=-\dem \ , & s_{17,58,61}=\dem \ , \\
s_{17,62,67}=-\dem \ , & s_{17,63,68}=-\dem \ , & s_{17,64,65}=-\dem \ , & s_{17,66,69}=\dem \ , \\
s_{17,72,77}=-1 \ , & s_{18,54,61}=\dem \ , & s_{18,55,57}=\dem \ , & s_{18,56,60}=-\dem \ , \\
s_{18,58,59}=\dem \ , & s_{18,62,69}=\dem \ , & s_{18,63,65}=\dem \ , & s_{18,64,68}=-\dem \ , \\
s_{18,66,67}=\dem \ , & s_{18,73,77}=-1 \ , & s_{19,54,58}=-\dem \ , & s_{19,55,56}=-\dem \ , \\
s_{19,57,60}=-\dem \ , & s_{19,59,61}=-\dem \ , & s_{19,62,66}=-\dem \ , & s_{19,63,64}=-\dem \ , \\
s_{19,65,68}=-\dem \ , & s_{19,67,69}=-\dem \ , & s_{19,74,77}=-1 \ , & s_{20,54,57}=\dem \ , \ea \farr \arr \bea{cccc}
s_{20,55,61}=-\dem \ , & s_{20,56,59}=-\dem \ , & s_{20,58,60}=-\dem \ , & s_{20,62,65}=\dem \ , \\
s_{20,63,69}=-\dem \ , & s_{20,64,67}=-\dem \ , & s_{20,66,68}=-\dem \ , & s_{20,75,77}=-1 \ , \\
s_{21,54,55}=\dem \ , & s_{21,56,58}=\dem \ , & s_{21,57,61}=\dem \ , & s_{21,59,60}=-\dem \ , \\
s_{21,62,63}=\dem \ , & s_{21,64,66}=\dem \ , & s_{21,65,69}=\dem \ , & s_{21,67,68}=-\dem \ , \\
s_{21,76,77}=-1 \ , & s_{22,53,61}=\dem \ , & s_{22,61,70}=-\frac {\sqrt 3}2 \ , & s_{22,62,78}=-\dem \ , \\
s_{22,63,74}=-\dem \ , & s_{22,64,77}=-\dem \ , & s_{22,65,72}=\dem \ , & s_{22,66,76}=-\dem \ , \\
s_{22,67,75}=\dem \ , & s_{22,68,73}=\dem \ , & s_{22,69,71}=\dem \ , & s_{23,53,57}=-\dem \ , \\
s_{23,57,70}=-\frac {\sqrt 3}2 \ , & s_{23,62,74}=\dem \ , & s_{23,63,78}=-\dem \ , & s_{23,64,76}=-\dem \ , \\
s_{23,65,71}=-\dem \ , & s_{23,66,77}=\dem \ , & s_{23,67,73}=\dem \ , & s_{23,68,75}=-\dem \ , \\
s_{23,69,72}=\dem \ , & s_{24,53,60}=-\dem \ , & s_{24,60,70}=-\frac {\sqrt 3}2 \ , & s_{24,62,77}=\dem \ , \\
s_{24,63,76}=\dem \ , & s_{24,64,78}=-\dem \ , & s_{24,65,75}=\dem \ , & s_{24,66,74}=-\dem \ , \\
s_{24,67,72}=-\dem \ , & s_{24,68,71}=-\dem \ , & s_{24,69,73}=\dem \ , & s_{25,53,55}=\dem \ , \\
s_{25,55,70}=\frac {\sqrt 3}2 \ , & s_{25,62,72}=-\dem \ , & s_{25,63,71}=\dem \ , & s_{25,64,75}=-\dem \ , \\
s_{25,65,78}=-\dem \ , & s_{25,66,73}=\dem \ , & s_{25,67,77}=-\dem \ , & s_{25,68,76}=\dem \ , \\
s_{25,68,74}=\dem \ , & s_{26,53,59}=-\dem \ , & s_{26,59,70}=-\frac {\sqrt 3}2 \ , & s_{26,62,76}=\dem \ , \\
s_{26,63,77}=-\dem \ , & s_{26,64,74}=\dem \ , & s_{26,65,73}=-\dem \ , & s_{26,66,78}=-\dem \ , \\
s_{26,67,71}=-\dem \ , & s_{26,68,72}=\dem \ , & s_{26,69,75}=\dem \ , & s_{27,53,58}=\dem \ , \\
s_{27,58,70}=\frac {\sqrt 3}2 \ , & s_{27,62,75}=-\dem \ , & s_{27,63,73}=-\dem \ , & s_{27,64,72}=\dem \ , \\
s_{27,64,77}=\dem \ , & s_{27,66,71}=\dem \ , & s_{27,67,78}=-\dem \ , & s_{27,68,74}=-\dem \ , \\
s_{27,69,76}=\dem \ , & s_{28,53,56}=\dem \ , & s_{28,56,70}=\frac {\sqrt 3}2 \ , & s_{28,62,73}=-\dem \ , \\
s_{28,63,75}=\dem \ , & s_{28,64,71}=\dem \ , & s_{28,65,76}=-\dem \ , & s_{28,66,72}=-\dem \ , \\
s_{28,67,74}=\dem \ , & s_{28,68,78}=-\dem \ , & s_{28,69,77}=\dem \ , & s_{29,53,54}=-\dem \ , \\
s_{29,54,70}=-\frac {\sqrt 3}2 \ , & s_{29,62,71}=-\dem \ , & s_{29,63,72}=-\dem \ , & s_{29,64,73}=-\dem \ , \\
s_{29,65,74}=-\dem \ , & s_{29,66,75}=-\dem \ , & s_{29,67,76}=-\dem \ , & s_{29,68,77}=-\dem \ , \\
s_{29,69,78}=-\dem \ , & s_{30,54,61}=\dem \ , & s_{30,55,57}=-\dem \ , & s_{30,56,60}=-\dem \ , \\
s_{30,58,59}=-\dem \ , & s_{30,62,69}=-\dem \ , & s_{30,63,65}=\dem \ , & s_{30,64,68}=\dem \ , \\
s_{30,66,67}=\dem \ , & s_{30,71,78}=-1 \ , & s_{31,54,57}=-\dem \ , & s_{31,55,61}=-\dem \ , \\
s_{31,56,59}=\dem \ , & s_{31,58,60}=-\dem \ , & s_{31,62,65}=-\dem \ , & s_{31,63,69}=-\dem \ , \\
s_{31,64,67}=\dem \ , & s_{31,66,68}=-\dem \ , & s_{31,72,78}=-1 \ , & s_{32,54,60}=-\dem \ , \\
s_{32,55,59}=-\dem \ , & s_{32,56,61}=-\dem \ , & s_{52,57,58}=-\dem \ , & s_{32,62,68}=-\dem \ , \\
s_{32,63,67}=-\dem \ , & s_{32,64,69}=-\dem \ , & s_{32,65,66}=-\dem \ , & s_{32,73,78}=-1 \ , \\
s_{33,54,55}=\dem \ , & s_{33,56,58}=\dem \ , & s_{33,57,61}=-\dem \ , & s_{33,59,60}=\dem \ , \\
s_{33,62,63}=\dem \ , & s_{33,64,66}=\dem \ , & s_{33,65,69}=-\dem \ , & s_{33,67,68}=\dem \ , \\
s_{33,74,78}=-1 \ , & s_{34,54,59}=-\dem \ , & s_{34,55,60}=\dem \ , & s_{34,56,57}=-\dem \ , \\
s_{34,58,61}=-\dem \ , & s_{34,62,67}=-\dem \ , & s_{34,63,68}=\dem \ , & s_{34,64,65}=-\dem \ , \\
s_{34,66,69}=-\dem \ , & s_{34,75,78}=-1 \ , & s_{35,54,58}=\dem \ , & s_{35,55,56}=\dem \ , \\
s_{35,57,60}=-\dem \ , & s_{35,59,61}=-\dem \ , & s_{35,62,66}=\dem \ , & s_{35,63,64}=\dem \ , \\
s_{35,65,68}=-\dem \ , & s_{35,67,69}=-\dem \ , & s_{35,76,78}=-1 \ , & s_{36,54,56}=\dem \ , \\
s_{36,55,58}=-\dem \ , & s_{36,57,59}=\dem \ , & s_{36,60,61}=-\dem \ , & s_{36,62,64}=\dem \ , \\
s_{36,63,66}=-\dem \ , & s_{36,65,67}=\dem \ , & s_{36,68,69}=-\dem \ , & s_{36,77,78}=-1 \ , \\
s_{37,53,62}=-1 \ , & s_{37,54,71}=-\dem \ , & s_{37,55,72}=\dem \ , & s_{37,56,73}=\dem \ , \\
s_{37,57,74}=\dem \ , & s_{37,58,75}=\dem \ , & s_{37,59,76}=\dem \ , & s_{37,60,77}=\dem \ , \\
s_{37,61,78}=\dem \ , & s_{38,53,63}=-1 \ , & s_{38,54,72}=-\dem \ , & s_{38,55,71}=-\dem \ , \\
s_{38,56,75}=-\dem \ , & s_{38,57,78}=-\dem \ , & s_{38,58,73}=\dem \ , & s_{38,59,77}=-\dem \ , \\
s_{38,60,76}=\dem \ , & s_{38,61,74}=\dem \ , & s_{39,53,64}=-1 \ , & s_{39,54,73}=-\dem \ , \ea \farr \arr \bea{cccc}
s_{39,55,75}=\dem \ , & s_{39,56,71}=-\dem \ , & s_{39,57,76}=-\dem \ , & s_{39,58,72}=-\dem \ , \\
s_{39,59,74}=\dem \ , & s_{39,60,78}=-\dem \ , & s_{39,61,77}=\dem \ , & s_{40,53,65}=-1 \ , \\
s_{40,54,74}=-\dem \ , & s_{40,55,78}=\dem \ , & s_{40,56,76}=\dem \ , & s_{40,57,71}=-\dem \ , \\
s_{40,58,77}=-\dem \ , & s_{40,59,73}=-\dem \ , & s_{40,60,75}=\dem \ , & s_{40,61,72}=-\dem \ , \\
s_{41,53,66}=-1 \ , & s_{41,54,75}=-\dem \ , & s_{41,55,73}=-\dem \ , & s_{41,56,72}=\dem \ , \\
s_{41,57,77}=\dem \ , & s_{41,58,71}=-\dem \ , & s_{41,59,78}=-\dem \ , & s_{41,60,74}=-\dem \ , \\
s_{41,61,76}=\dem \ , & s_{42,53,67}=-1 \ , & s_{42,54,76}=-\dem \ , & s_{42,55,77}=\dem \ , \\
s_{42,56,74}=-\dem \ , & s_{42,57,73}=\dem \ , & s_{42,58,78}=\dem \ , & s_{42,59,71}=-\dem \ , \\
s_{42,60,72}=-\dem \ , & s_{42,61,75}=-\dem \ , & s_{43,53,68}=-1 \ , & s_{43,54,77}=-\dem \ , \\
s_{43,55,76}=-\dem \ , & s_{43,56,78}=\dem \ , & s_{43,57,75}=-\dem \ , & s_{43,58,74}=\dem \ , \\
s_{43,59,72}=\dem \ , & s_{43,60,71}=-\dem \ , & s_{43,61,73}=-\dem \ , & s_{44,53,69}=-1 \ , \\
s_{44,54,78}=-\dem \ , & s_{44,55,74}=-\dem \ , & s_{44,56,77}=-\dem \ , & s_{44,57,72}=\dem \ , \\
s_{44,58,76}=-\dem \ , & s_{44,59,75}=\dem \ , & s_{44,60,73}=\dem \ , & s_{44,61,71}=-\dem \ , \\
s_{45,53,71}=-\dem \ , & s_{45,54,62}=-\dem \ , & s_{45,55,63}=-\dem \ , & s_{45,56,64}=-\dem \ , \\
s_{45,57,65}=-\dem \ , & s_{45,58,66}=-\dem \ , & s_{45,59,67}=-\dem \ , & s_{45,60,68}=-\dem \ , \\
s_{45,61,69}=-\dem \ , & s_{45,70,71}=-\frac {\sqrt 3}2 \ , & s_{46,53,72}=-\dem \ , & s_{46,54,63}=-\dem \ , \\
s_{46,55,62}=\dem \ , & s_{46,56,66}=\dem \ , & s_{46,57,69}=\dem \ , & s_{46,58,64}=-\dem \ , \\
s_{46,59,68}=\dem \ , & s_{46,60,67}=-\dem \ , & s_{46,61,65}=-\dem \ , & s_{46,70,72}=-\frac {\sqrt 3}2 \ , \\
s_{47,53,73}=-\dem \ , & s_{47,54,64}=-\dem \ , & s_{47,55,66}=-\dem \ , & s_{47,56,62}=\dem \ , \\
s_{47,57,67}=\dem \ , & s_{47,58,63}=\dem \ , & s_{47,59,65}=-\dem \ , & s_{47,60,69}=\dem \ , \\
s_{47,61,68}=-\dem \ , & s_{47,70,73}=-\frac {\sqrt 3}2 \ , & s_{48,53,74}=-\dem \ , & s_{48,54,65}=-\dem \ , \\
s_{48,55,69}=-\dem \ , & s_{48,56,67}=-\dem \ , & s_{48,57,62}=-\dem \ , & s_{48,58,68}=\dem \ , \\
s_{48,59,64}=\dem \ , & s_{48,60,66}=-\dem \ , & s_{48,61,63}=\dem \ , & s_{48,70,74}=-\frac {\sqrt 3}2 \ , \\
s_{49,53,75}=-\dem \ , & s_{49,54,66}=-\dem \ , & s_{49,55,64}=\dem \ , & s_{49,56,63}=-\dem \ , \\
s_{49,57,68}=-\dem \ , & s_{49,58,62}=\dem \ , & s_{49,59,69}=\dem \ , & s_{49,60,65}=\dem \ , \\
s_{49,61,67}=-\dem \ , & s_{49,70,75}=-\frac {\sqrt 3}2 \ , & s_{50,53,76}=-\dem \ , & s_{50,54,67}=-\dem \ , \\
s_{50,55,68}=-\dem \ , & s_{50,56,65}=\dem \ , & s_{50,57,64}=-\dem \ , & s_{50,58,69}=-\dem \ , \\
s_{50,59,62}=\dem \ , & s_{50,60,63}=\dem \ , & s_{50,61,66}=\dem \ , & s_{50,70,76}=-\frac {\sqrt 3}2 \ , \\
s_{51,53,77}=-\dem \ , & s_{51,54,68}=-\dem \ , & s_{51,55,67}=\dem \ , & s_{51,56,69}=-\dem \ , \\
s_{51,57,66}=\dem \ , & s_{51,58,65}=-\dem \ , & s_{51,59,63}=-\dem \ , & s_{51,60,62}=\dem \ , \\
s_{51,61,64}=\dem \ , & s_{51,70,77}=-\frac {\sqrt 3}2 \ , & s_{52,53,78}=-\dem \ , & s_{52,54,69}=-\dem \ , \\
s_{52,55,65}=\dem \ , & s_{52,56,68}=-\dem \ , & s_{52,57,63}=-\dem \ , & s_{52,58,67}=\dem \ , \\
s_{52,59,66}=-\dem \ , & s_{52,60,64}=-\dem \ , & s_{52,61,62}=\dem \ , & s_{52,70,78}=-\frac {\sqrt 3}2 \ .
\ea \farr

\section{The matrices.}
Here we write the matrices $c_i$, $i=53,\ldots,78$.
{\tiny
\eqn
c_{53}=\lp
\begin{array}{ccccccccccccccccccccccccccc}
-\iem,\!\!\!\!\!\!\!\! &0,\!\!\!\!\!\!\!\! &0,\!\!\!\!\!\!\!\! &0,\!\!\!\!\!\!\!\! &0,\!\!\!\!\!\!\!\! &0,\!\!\!\!\!\!\!\! &0,\!\!\!\!\!\!\!\! &0,\!\!\!\!\!\!\!\! &0,\!\!\!\!\!\!\!\! &0,\!\!\!\!\!\!\!\! &0,\!\!\!\!\!\!\!\! &0,\!\!\!\!\!\!\!\! &0,\!\!\!\!\!\!\!\! &0,\!\!\!\!\!\!\!\! &0,\!\!\!\!\!\!\!\! &0,\!\!\!\!\!\!\!\! &0,\!\!\!\!\!\!\!\! &-\frac i{2\sqrt 3},\!\!\!\!\!\!\!\! &0,\!\!\!\!\!\!\!\! &0,\!\!\!\!\!\!\!\! &0,\!\!\!\!\!\!\!\! &0,\!\!\!\!\!\!\!\! &0,\!\!\!\!\!\!\!\! &0,\!\!\!\!\!\!\!\! &0,\!\!\!\!\!\!\!\! &0,\!\!\!\!\!\!\!\! &-\frac i{\sqrt 6} \\
0,\!\!\!\!\!\!\!\! &-\iem,\!\!\!\!\!\!\!\! &0,\!\!\!\!\!\!\!\! &0,\!\!\!\!\!\!\!\! &0,\!\!\!\!\!\!\!\! &0,\!\!\!\!\!\!\!\! &0,\!\!\!\!\!\!\!\! &0,\!\!\!\!\!\!\!\! &0,\!\!\!\!\!\!\!\! &0,\!\!\!\!\!\!\!\! &0,\!\!\!\!\!\!\!\! &0,\!\!\!\!\!\!\!\! &0,\!\!\!\!\!\!\!\! &0,\!\!\!\!\!\!\!\! &0,\!\!\!\!\!\!\!\! &0,\!\!\!\!\!\!\!\! &0,\!\!\!\!\!\!\!\! &0,\!\!\!\!\!\!\!\! &0,\!\!\!\!\!\!\!\! &0,\!\!\!\!\!\!\!\! &0,\!\!\!\!\!\!\!\! &0,\!\!\!\!\!\!\!\! &0,\!\!\!\!\!\!\!\! &0,\!\!\!\!\!\!\!\! &0,\!\!\!\!\!\!\!\! &0,\!\!\!\!\!\!\!\! &0 \\
0,\!\!\!\!\!\!\!\! &0,\!\!\!\!\!\!\!\! &-\iem,\!\!\!\!\!\!\!\! &0,\!\!\!\!\!\!\!\! &0,\!\!\!\!\!\!\!\! &0,\!\!\!\!\!\!\!\! &0,\!\!\!\!\!\!\!\! &0,\!\!\!\!\!\!\!\! &0,\!\!\!\!\!\!\!\! &0,\!\!\!\!\!\!\!\! &0,\!\!\!\!\!\!\!\! &0,\!\!\!\!\!\!\!\! &0,\!\!\!\!\!\!\!\! &0,\!\!\!\!\!\!\!\! &0,\!\!\!\!\!\!\!\! &0,\!\!\!\!\!\!\!\! &0,\!\!\!\!\!\!\!\! &0,\!\!\!\!\!\!\!\! &0,\!\!\!\!\!\!\!\! &0,\!\!\!\!\!\!\!\! &0,\!\!\!\!\!\!\!\! &0,\!\!\!\!\!\!\!\! &0,\!\!\!\!\!\!\!\! &0,\!\!\!\!\!\!\!\! &0,\!\!\!\!\!\!\!\! &0,\!\!\!\!\!\!\!\! &0 \\
\end{array}
\rp
\feqn
\eqn
c_{54}=\lp
\begin{array}{ccccccccccccccccccccccccccc}
0,\!\!\!\!\!\!\!\! &0,\!\!\!\!\!\!\!\! &0,\!\!\!\!\!\!\!\! &0,\!\!\!\!\!\!\!\! &0,\!\!\!\!\!\!\!\! &0,\!\!\!\!\!\!\!\! &0,\!\!\!\!\!\!\!\! &0,\!\!\!\!\!\!\!\! &0,\!\!\!\!\!\!\!\! &0,\!\!\!\!\!\!\!\! &0,\!\!\!\!\!\!\!\! &0,\!\!\!\!\!\!\!\! &0,\!\!\!\!\!\!\!\! &0,\!\!\!\!\!\!\!\! &0,\!\!\!\!\!\!\!\! &0,\!\!\!\!\!\!\!\! &0,\!\!\!\!\!\!\!\! &0,\!\!\!\!\!\!\!\! &0,\!\!\!\!\!\!\!\! &0,\!\!\!\!\!\!\!\! &0,\!\!\!\!\!\!\!\! &0,\!\!\!\!\!\!\!\! &0,\!\!\!\!\!\!\!\! &0,\!\!\!\!\!\!\!\! &0,\!\!\!\!\!\!\!\! &0,\!\!\!\!\!\!\!\! &0 \\
0,\!\!\!\!\!\!\!\! &0,\!\!\!\!\!\!\!\! &0,\!\!\!\!\!\!\!\! &0,\!\!\!\!\!\!\!\! &0,\!\!\!\!\!\!\!\! &0,\!\!\!\!\!\!\!\! &0,\!\!\!\!\!\!\!\! &0,\!\!\!\!\!\!\!\! &0,\!\!\!\!\!\!\!\! &0,\!\!\!\!\!\!\!\! &0,\!\!\!\!\!\!\!\! &0,\!\!\!\!\!\!\!\! &0,\!\!\!\!\!\!\!\! &0,\!\!\!\!\!\!\!\! &0,\!\!\!\!\!\!\!\! &0,\!\!\!\!\!\!\!\! &0,\!\!\!\!\!\!\!\! &-\frac i{\sqrt 6},\!\!\!\!\!\!\!\! &0,\!\!\!\!\!\!\!\! &0,\!\!\!\!\!\!\!\! &0,\!\!\!\!\!\!\!\! &0,\!\!\!\!\!\!\!\! &0,\!\!\!\!\!\!\!\! &0,\!\!\!\!\!\!\!\! &0,\!\!\!\!\!\!\!\! &0,\!\!\!\!\!\!\!\! &-\frac i{\sqrt 3} \\
0,\!\!\!\!\!\!\!\! &0,\!\!\!\!\!\!\!\! &0,\!\!\!\!\!\!\!\! &0,\!\!\!\!\!\!\!\! &0,\!\!\!\!\!\!\!\! &0,\!\!\!\!\!\!\!\! &0,\!\!\!\!\!\!\!\! &0,\!\!\!\!\!\!\!\! &0,\!\!\!\!\!\!\!\! &0,\!\!\!\!\!\!\!\! &0,\!\!\!\!\!\!\!\! &0,\!\!\!\!\!\!\!\! &0,\!\!\!\!\!\!\!\! &0,\!\!\!\!\!\!\!\! &0,\!\!\!\!\!\!\!\! &0,\!\!\!\!\!\!\!\! &0,\!\!\!\!\!\!\!\! &0,\!\!\!\!\!\!\!\! &0,\!\!\!\!\!\!\!\! &0,\!\!\!\!\!\!\!\! &0,\!\!\!\!\!\!\!\! &0,\!\!\!\!\!\!\!\! &0,\!\!\!\!\!\!\!\! &0,\!\!\!\!\!\!\!\! &0,\!\!\!\!\!\!\!\! &0,\!\!\!\!\!\!\!\! &0 \\
\end{array}
\rp
\feqn
\eqn
c_{55}=\lp
\begin{array}{ccccccccccccccccccccccccccc}
0,\!\!\!\!\!\!\!\! &0,\!\!\!\!\!\!\!\! &0,\!\!\!\!\!\!\!\! &0,\!\!\!\!\!\!\!\! &0,\!\!\!\!\!\!\!\! &0,\!\!\!\!\!\!\!\! &0,\!\!\!\!\!\!\!\! &0,\!\!\!\!\!\!\!\! &0,\!\!\!\!\!\!\!\! &0,\!\!\!\!\!\!\!\! &0,\!\!\!\!\!\!\!\! &0,\!\!\!\!\!\!\!\! &0,\!\!\!\!\!\!\!\! &0,\!\!\!\!\!\!\!\! &0,\!\!\!\!\!\!\!\! &0,\!\!\!\!\!\!\!\! &0,\!\!\!\!\!\!\!\! &0,\!\!\!\!\!\!\!\! &0,\!\!\!\!\!\!\!\! &0,\!\!\!\!\!\!\!\! &0,\!\!\!\!\!\!\!\! &0,\!\!\!\!\!\!\!\! &0,\!\!\!\!\!\!\!\! &0,\!\!\!\!\!\!\!\! &0,\!\!\!\!\!\!\!\! &0,\!\!\!\!\!\!\!\! &0 \\
0,\!\!\!\!\!\!\!\! &0,\!\!\!\!\!\!\!\! &0,\!\!\!\!\!\!\!\! &0,\!\!\!\!\!\!\!\! &0,\!\!\!\!\!\!\!\! &0,\!\!\!\!\!\!\!\! &0,\!\!\!\!\!\!\!\! &0,\!\!\!\!\!\!\!\! &0,\!\!\!\!\!\!\!\! &0,\!\!\!\!\!\!\!\! &0,\!\!\!\!\!\!\!\! &0,\!\!\!\!\!\!\!\! &0,\!\!\!\!\!\!\!\! &0,\!\!\!\!\!\!\!\! &0,\!\!\!\!\!\!\!\! &0,\!\!\!\!\!\!\!\! &0,\!\!\!\!\!\!\!\! &0,\!\!\!\!\!\!\!\! &0,\!\!\!\!\!\!\!\! &0,\!\!\!\!\!\!\!\! &0,\!\!\!\!\!\!\!\! &0,\!\!\!\!\!\!\!\! &0,\!\!\!\!\!\!\!\! &0,\!\!\!\!\!\!\!\! &0,\!\!\!\!\!\!\!\! &0,\!\!\!\!\!\!\!\! &0 \\
0,\!\!\!\!\!\!\!\! &0,\!\!\!\!\!\!\!\! &0,\!\!\!\!\!\!\!\! &0,\!\!\!\!\!\!\!\! &0,\!\!\!\!\!\!\!\! &0,\!\!\!\!\!\!\!\! &0,\!\!\!\!\!\!\!\! &0,\!\!\!\!\!\!\!\! &0,\!\!\!\!\!\!\!\! &0,\!\!\!\!\!\!\!\! &0,\!\!\!\!\!\!\!\! &0,\!\!\!\!\!\!\!\! &0,\!\!\!\!\!\!\!\! &0,\!\!\!\!\!\!\!\! &0,\!\!\!\!\!\!\!\! &0,\!\!\!\!\!\!\!\! &0,\!\!\!\!\!\!\!\! &-\frac i{\sqrt 6},\!\!\!\!\!\!\!\! &0,\!\!\!\!\!\!\!\! &0,\!\!\!\!\!\!\!\! &0,\!\!\!\!\!\!\!\! &0,\!\!\!\!\!\!\!\! &0,\!\!\!\!\!\!\!\! &0,\!\!\!\!\!\!\!\! &0,\!\!\!\!\!\!\!\! &0,\!\!\!\!\!\!\!\! &-\frac i{\sqrt 3} \\
\end{array}
\rp
\feqn
\eqn
c_{56}=\lp
\begin{array}{ccccccccccccccccccccccccccc}
0,\!\!\!\!\!\!\!\! &0,\!\!\!\!\!\!\!\! &0,\!\!\!\!\!\!\!\! &0,\!\!\!\!\!\!\!\! &0,\!\!\!\!\!\!\!\! &0,\!\!\!\!\!\!\!\! &0,\!\!\!\!\!\!\!\! &0,\!\!\!\!\!\!\!\! &0,\!\!\!\!\!\!\!\! &0,\!\!\!\!\!\!\!\! &0,\!\!\!\!\!\!\!\! &0,\!\!\!\!\!\!\!\! &0,\!\!\!\!\!\!\!\! &0,\!\!\!\!\!\!\!\! &0,\!\!\!\!\!\!\!\! &0,\!\!\!\!\!\!\!\! &0,\!\!\!\!\!\!\!\! &0,\!\!\!\!\!\!\!\! &0,\!\!\!\!\!\!\!\! &0,\!\!\!\!\!\!\!\! &0,\!\!\!\!\!\!\!\! &0,\!\!\!\!\!\!\!\! &0,\!\!\!\!\!\!\!\! &0,\!\!\!\!\!\!\!\! &0,\!\!\!\!\!\!\!\! &0,\!\!\!\!\!\!\!\! &0 \\
0,\!\!\!\!\!\!\!\! &0,\!\!\!\!\!\!\!\! &0,\!\!\!\!\!\!\!\! &0,\!\!\!\!\!\!\!\! &0,\!\!\!\!\!\!\!\! &0,\!\!\!\!\!\!\!\! &0,\!\!\!\!\!\!\!\! &0,\!\!\!\!\!\!\!\! &0,\!\!\!\!\!\!\!\! &0,\!\!\!\!\!\!\!\! &0,\!\!\!\!\!\!\!\! &0,\!\!\!\!\!\!\!\! &0,\!\!\!\!\!\!\!\! &0,\!\!\!\!\!\!\!\! &0,\!\!\!\!\!\!\!\! &0,\!\!\!\!\!\!\!\! &0,\!\!\!\!\!\!\!\! &0,\!\!\!\!\!\!\!\! &0,\!\!\!\!\!\!\!\! &0,\!\!\!\!\!\!\!\! &0,\!\!\!\!\!\!\!\! &0,\!\!\!\!\!\!\!\! &0,\!\!\!\!\!\!\!\! &0,\!\!\!\!\!\!\!\! &0,\!\!\!\!\!\!\!\! &0,\!\!\!\!\!\!\!\! &0 \\
0,\!\!\!\!\!\!\!\! &0,\!\!\!\!\!\!\!\! &0,\!\!\!\!\!\!\!\! &0,\!\!\!\!\!\!\!\! &0,\!\!\!\!\!\!\!\! &0,\!\!\!\!\!\!\!\! &0,\!\!\!\!\!\!\!\! &0,\!\!\!\!\!\!\!\! &0,\!\!\!\!\!\!\!\! &0,\!\!\!\!\!\!\!\! &0,\!\!\!\!\!\!\!\! &0,\!\!\!\!\!\!\!\! &0,\!\!\!\!\!\!\!\! &0,\!\!\!\!\!\!\!\! &0,\!\!\!\!\!\!\!\! &0,\!\!\!\!\!\!\!\! &0,\!\!\!\!\!\!\!\! &0,\!\!\!\!\!\!\!\! &0,\!\!\!\!\!\!\!\! &0,\!\!\!\!\!\!\!\! &0,\!\!\!\!\!\!\!\! &0,\!\!\!\!\!\!\!\! &0,\!\!\!\!\!\!\!\! &0,\!\!\!\!\!\!\!\! &0,\!\!\!\!\!\!\!\! &0,\!\!\!\!\!\!\!\! &0 \\
\end{array}
\rp
\feqn
\eqn
c_{57}=\lp
\begin{array}{ccccccccccccccccccccccccccc}
0,\!\!\!\!\!\!\!\! &0,\!\!\!\!\!\!\!\! &0,\!\!\!\!\!\!\!\! &0,\!\!\!\!\!\!\!\! &0,\!\!\!\!\!\!\!\! &0,\!\!\!\!\!\!\!\! &0,\!\!\!\!\!\!\!\! &0,\!\!\!\!\!\!\!\! &0,\!\!\!\!\!\!\!\! &0,\!\!\!\!\!\!\!\! &0,\!\!\!\!\!\!\!\! &0,\!\!\!\!\!\!\!\! &0,\!\!\!\!\!\!\!\! &0,\!\!\!\!\!\!\!\! &0,\!\!\!\!\!\!\!\! &0,\!\!\!\!\!\!\!\! &0,\!\!\!\!\!\!\!\! &0,\!\!\!\!\!\!\!\! &0,\!\!\!\!\!\!\!\! &0,\!\!\!\!\!\!\!\! &0,\!\!\!\!\!\!\!\! &0,\!\!\!\!\!\!\!\! &0,\!\!\!\!\!\!\!\! &0,\!\!\!\!\!\!\!\! &0,\!\!\!\!\!\!\!\! &0,\!\!\!\!\!\!\!\! &0 \\
0,\!\!\!\!\!\!\!\! &0,\!\!\!\!\!\!\!\! &0,\!\!\!\!\!\!\!\! &0,\!\!\!\!\!\!\!\! &0,\!\!\!\!\!\!\!\! &0,\!\!\!\!\!\!\!\! &0,\!\!\!\!\!\!\!\! &0,\!\!\!\!\!\!\!\! &0,\!\!\!\!\!\!\!\! &0,\!\!\!\!\!\!\!\! &0,\!\!\!\!\!\!\!\! &0,\!\!\!\!\!\!\!\! &0,\!\!\!\!\!\!\!\! &0,\!\!\!\!\!\!\!\! &0,\!\!\!\!\!\!\!\! &0,\!\!\!\!\!\!\!\! &0,\!\!\!\!\!\!\!\! &0,\!\!\!\!\!\!\!\! &0,\!\!\!\!\!\!\!\! &0,\!\!\!\!\!\!\!\! &0,\!\!\!\!\!\!\!\! &0,\!\!\!\!\!\!\!\! &0,\!\!\!\!\!\!\!\! &0,\!\!\!\!\!\!\!\! &0,\!\!\!\!\!\!\!\! &0,\!\!\!\!\!\!\!\! &0 \\
0,\!\!\!\!\!\!\!\! &0,\!\!\!\!\!\!\!\! &0,\!\!\!\!\!\!\!\! &0,\!\!\!\!\!\!\!\! &0,\!\!\!\!\!\!\!\! &0,\!\!\!\!\!\!\!\! &0,\!\!\!\!\!\!\!\! &0,\!\!\!\!\!\!\!\! &0,\!\!\!\!\!\!\!\! &0,\!\!\!\!\!\!\!\! &0,\!\!\!\!\!\!\!\! &0,\!\!\!\!\!\!\!\! &0,\!\!\!\!\!\!\!\! &0,\!\!\!\!\!\!\!\! &0,\!\!\!\!\!\!\!\! &0,\!\!\!\!\!\!\!\! &0,\!\!\!\!\!\!\!\! &0,\!\!\!\!\!\!\!\! &0,\!\!\!\!\!\!\!\! &0,\!\!\!\!\!\!\!\! &0,\!\!\!\!\!\!\!\! &0,\!\!\!\!\!\!\!\! &0,\!\!\!\!\!\!\!\! &0,\!\!\!\!\!\!\!\! &0,\!\!\!\!\!\!\!\! &0,\!\!\!\!\!\!\!\! &0 \\
\end{array}
\rp
\feqn
\eqn
c_{58}=\lp
\begin{array}{ccccccccccccccccccccccccccc}
0,\!\!\!\!\!\!\!\! &0,\!\!\!\!\!\!\!\! &0,\!\!\!\!\!\!\!\! &0,\!\!\!\!\!\!\!\! &0,\!\!\!\!\!\!\!\! &0,\!\!\!\!\!\!\!\! &0,\!\!\!\!\!\!\!\! &0,\!\!\!\!\!\!\!\! &0,\!\!\!\!\!\!\!\! &0,\!\!\!\!\!\!\!\! &0,\!\!\!\!\!\!\!\! &0,\!\!\!\!\!\!\!\! &0,\!\!\!\!\!\!\!\! &0,\!\!\!\!\!\!\!\! &0,\!\!\!\!\!\!\!\! &0,\!\!\!\!\!\!\!\! &0,\!\!\!\!\!\!\!\! &0,\!\!\!\!\!\!\!\! &0,\!\!\!\!\!\!\!\! &0,\!\!\!\!\!\!\!\! &0,\!\!\!\!\!\!\!\! &0,\!\!\!\!\!\!\!\! &0,\!\!\!\!\!\!\!\! &0,\!\!\!\!\!\!\!\! &0,\!\!\!\!\!\!\!\! &0,\!\!\!\!\!\!\!\! &0 \\
0,\!\!\!\!\!\!\!\! &0,\!\!\!\!\!\!\!\! &0,\!\!\!\!\!\!\!\! &0,\!\!\!\!\!\!\!\! &0,\!\!\!\!\!\!\!\! &0,\!\!\!\!\!\!\!\! &0,\!\!\!\!\!\!\!\! &0,\!\!\!\!\!\!\!\! &0,\!\!\!\!\!\!\!\! &0,\!\!\!\!\!\!\!\! &0,\!\!\!\!\!\!\!\! &0,\!\!\!\!\!\!\!\! &0,\!\!\!\!\!\!\!\! &0,\!\!\!\!\!\!\!\! &0,\!\!\!\!\!\!\!\! &0,\!\!\!\!\!\!\!\! &0,\!\!\!\!\!\!\!\! &0,\!\!\!\!\!\!\!\! &0,\!\!\!\!\!\!\!\! &0,\!\!\!\!\!\!\!\! &0,\!\!\!\!\!\!\!\! &0,\!\!\!\!\!\!\!\! &0,\!\!\!\!\!\!\!\! &0,\!\!\!\!\!\!\!\! &0,\!\!\!\!\!\!\!\! &0,\!\!\!\!\!\!\!\! &0 \\
0,\!\!\!\!\!\!\!\! &0,\!\!\!\!\!\!\!\! &0,\!\!\!\!\!\!\!\! &0,\!\!\!\!\!\!\!\! &0,\!\!\!\!\!\!\!\! &0,\!\!\!\!\!\!\!\! &0,\!\!\!\!\!\!\!\! &0,\!\!\!\!\!\!\!\! &0,\!\!\!\!\!\!\!\! &0,\!\!\!\!\!\!\!\! &0,\!\!\!\!\!\!\!\! &0,\!\!\!\!\!\!\!\! &0,\!\!\!\!\!\!\!\! &0,\!\!\!\!\!\!\!\! &0,\!\!\!\!\!\!\!\! &0,\!\!\!\!\!\!\!\! &0,\!\!\!\!\!\!\!\! &0,\!\!\!\!\!\!\!\! &0,\!\!\!\!\!\!\!\! &0,\!\!\!\!\!\!\!\! &0,\!\!\!\!\!\!\!\! &0,\!\!\!\!\!\!\!\! &0,\!\!\!\!\!\!\!\! &0,\!\!\!\!\!\!\!\! &0,\!\!\!\!\!\!\!\! &0,\!\!\!\!\!\!\!\! &0 \\
\end{array}
\rp
\feqn
\eqn
c_{59}=\lp
\begin{array}{ccccccccccccccccccccccccccc}
0,\!\!\!\!\!\!\!\! &0,\!\!\!\!\!\!\!\! &0,\!\!\!\!\!\!\!\! &0,\!\!\!\!\!\!\!\! &0,\!\!\!\!\!\!\!\! &0,\!\!\!\!\!\!\!\! &0,\!\!\!\!\!\!\!\! &0,\!\!\!\!\!\!\!\! &0,\!\!\!\!\!\!\!\! &0,\!\!\!\!\!\!\!\! &0,\!\!\!\!\!\!\!\! &0,\!\!\!\!\!\!\!\! &0,\!\!\!\!\!\!\!\! &0,\!\!\!\!\!\!\!\! &0,\!\!\!\!\!\!\!\! &0,\!\!\!\!\!\!\!\! &0,\!\!\!\!\!\!\!\! &0,\!\!\!\!\!\!\!\! &0,\!\!\!\!\!\!\!\! &0,\!\!\!\!\!\!\!\! &0,\!\!\!\!\!\!\!\! &0,\!\!\!\!\!\!\!\! &0,\!\!\!\!\!\!\!\! &0,\!\!\!\!\!\!\!\! &0,\!\!\!\!\!\!\!\! &0,\!\!\!\!\!\!\!\! &0 \\
0,\!\!\!\!\!\!\!\! &0,\!\!\!\!\!\!\!\! &0,\!\!\!\!\!\!\!\! &0,\!\!\!\!\!\!\!\! &0,\!\!\!\!\!\!\!\! &0,\!\!\!\!\!\!\!\! &0,\!\!\!\!\!\!\!\! &0,\!\!\!\!\!\!\!\! &0,\!\!\!\!\!\!\!\! &0,\!\!\!\!\!\!\!\! &0,\!\!\!\!\!\!\!\! &0,\!\!\!\!\!\!\!\! &0,\!\!\!\!\!\!\!\! &0,\!\!\!\!\!\!\!\! &0,\!\!\!\!\!\!\!\! &0,\!\!\!\!\!\!\!\! &0,\!\!\!\!\!\!\!\! &0,\!\!\!\!\!\!\!\! &0,\!\!\!\!\!\!\!\! &0,\!\!\!\!\!\!\!\! &0,\!\!\!\!\!\!\!\! &0,\!\!\!\!\!\!\!\! &0,\!\!\!\!\!\!\!\! &0,\!\!\!\!\!\!\!\! &0,\!\!\!\!\!\!\!\! &0,\!\!\!\!\!\!\!\! &0 \\
0,\!\!\!\!\!\!\!\! &0,\!\!\!\!\!\!\!\! &0,\!\!\!\!\!\!\!\! &0,\!\!\!\!\!\!\!\! &0,\!\!\!\!\!\!\!\! &0,\!\!\!\!\!\!\!\! &0,\!\!\!\!\!\!\!\! &0,\!\!\!\!\!\!\!\! &0,\!\!\!\!\!\!\!\! &0,\!\!\!\!\!\!\!\! &0,\!\!\!\!\!\!\!\! &0,\!\!\!\!\!\!\!\! &0,\!\!\!\!\!\!\!\! &0,\!\!\!\!\!\!\!\! &0,\!\!\!\!\!\!\!\! &0,\!\!\!\!\!\!\!\! &0,\!\!\!\!\!\!\!\! &0,\!\!\!\!\!\!\!\! &0,\!\!\!\!\!\!\!\! &0,\!\!\!\!\!\!\!\! &0,\!\!\!\!\!\!\!\! &0,\!\!\!\!\!\!\!\! &0,\!\!\!\!\!\!\!\! &0,\!\!\!\!\!\!\!\! &0,\!\!\!\!\!\!\!\! &0,\!\!\!\!\!\!\!\! &0 \\
\end{array}
\rp
\feqn
\eqn
c_{60}=\lp
\begin{array}{ccccccccccccccccccccccccccc}
0,\!\!\!\!\!\!\!\! &0,\!\!\!\!\!\!\!\! &0,\!\!\!\!\!\!\!\! &0,\!\!\!\!\!\!\!\! &0,\!\!\!\!\!\!\!\! &0,\!\!\!\!\!\!\!\! &0,\!\!\!\!\!\!\!\! &0,\!\!\!\!\!\!\!\! &0,\!\!\!\!\!\!\!\! &0,\!\!\!\!\!\!\!\! &0,\!\!\!\!\!\!\!\! &0,\!\!\!\!\!\!\!\! &0,\!\!\!\!\!\!\!\! &0,\!\!\!\!\!\!\!\! &0,\!\!\!\!\!\!\!\! &0,\!\!\!\!\!\!\!\! &0,\!\!\!\!\!\!\!\! &0,\!\!\!\!\!\!\!\! &0,\!\!\!\!\!\!\!\! &0,\!\!\!\!\!\!\!\! &0,\!\!\!\!\!\!\!\! &0,\!\!\!\!\!\!\!\! &0,\!\!\!\!\!\!\!\! &0,\!\!\!\!\!\!\!\! &0,\!\!\!\!\!\!\!\! &0,\!\!\!\!\!\!\!\! &0 \\
0,\!\!\!\!\!\!\!\! &0,\!\!\!\!\!\!\!\! &0,\!\!\!\!\!\!\!\! &0,\!\!\!\!\!\!\!\! &0,\!\!\!\!\!\!\!\! &0,\!\!\!\!\!\!\!\! &0,\!\!\!\!\!\!\!\! &0,\!\!\!\!\!\!\!\! &0,\!\!\!\!\!\!\!\! &0,\!\!\!\!\!\!\!\! &0,\!\!\!\!\!\!\!\! &0,\!\!\!\!\!\!\!\! &0,\!\!\!\!\!\!\!\! &0,\!\!\!\!\!\!\!\! &0,\!\!\!\!\!\!\!\! &0,\!\!\!\!\!\!\!\! &0,\!\!\!\!\!\!\!\! &0,\!\!\!\!\!\!\!\! &0,\!\!\!\!\!\!\!\! &0,\!\!\!\!\!\!\!\! &0,\!\!\!\!\!\!\!\! &0,\!\!\!\!\!\!\!\! &0,\!\!\!\!\!\!\!\! &0,\!\!\!\!\!\!\!\! &0,\!\!\!\!\!\!\!\! &0,\!\!\!\!\!\!\!\! &0 \\
0,\!\!\!\!\!\!\!\! &0,\!\!\!\!\!\!\!\! &0,\!\!\!\!\!\!\!\! &0,\!\!\!\!\!\!\!\! &0,\!\!\!\!\!\!\!\! &0,\!\!\!\!\!\!\!\! &0,\!\!\!\!\!\!\!\! &0,\!\!\!\!\!\!\!\! &0,\!\!\!\!\!\!\!\! &0,\!\!\!\!\!\!\!\! &0,\!\!\!\!\!\!\!\! &0,\!\!\!\!\!\!\!\! &0,\!\!\!\!\!\!\!\! &0,\!\!\!\!\!\!\!\! &0,\!\!\!\!\!\!\!\! &0,\!\!\!\!\!\!\!\! &0,\!\!\!\!\!\!\!\! &0,\!\!\!\!\!\!\!\! &0,\!\!\!\!\!\!\!\! &0,\!\!\!\!\!\!\!\! &0,\!\!\!\!\!\!\!\! &0,\!\!\!\!\!\!\!\! &0,\!\!\!\!\!\!\!\! &0,\!\!\!\!\!\!\!\! &0,\!\!\!\!\!\!\!\! &0,\!\!\!\!\!\!\!\! &0 \\
\end{array}
\rp
\feqn
\eqn
c_{61}=\lp
\begin{array}{ccccccccccccccccccccccccccc}
0,\!\!\!\!\!\!\!\! &0,\!\!\!\!\!\!\!\! &0,\!\!\!\!\!\!\!\! &0,\!\!\!\!\!\!\!\! &0,\!\!\!\!\!\!\!\! &0,\!\!\!\!\!\!\!\! &0,\!\!\!\!\!\!\!\! &0,\!\!\!\!\!\!\!\! &0,\!\!\!\!\!\!\!\! &0,\!\!\!\!\!\!\!\! &0,\!\!\!\!\!\!\!\! &0,\!\!\!\!\!\!\!\! &0,\!\!\!\!\!\!\!\! &0,\!\!\!\!\!\!\!\! &0,\!\!\!\!\!\!\!\! &0,\!\!\!\!\!\!\!\! &0,\!\!\!\!\!\!\!\! &0,\!\!\!\!\!\!\!\! &0,\!\!\!\!\!\!\!\! &0,\!\!\!\!\!\!\!\! &0,\!\!\!\!\!\!\!\! &0,\!\!\!\!\!\!\!\! &0,\!\!\!\!\!\!\!\! &0,\!\!\!\!\!\!\!\! &0,\!\!\!\!\!\!\!\! &0,\!\!\!\!\!\!\!\! &0 \\
0,\!\!\!\!\!\!\!\! &0,\!\!\!\!\!\!\!\! &0,\!\!\!\!\!\!\!\! &0,\!\!\!\!\!\!\!\! &0,\!\!\!\!\!\!\!\! &0,\!\!\!\!\!\!\!\! &0,\!\!\!\!\!\!\!\! &0,\!\!\!\!\!\!\!\! &0,\!\!\!\!\!\!\!\! &0,\!\!\!\!\!\!\!\! &0,\!\!\!\!\!\!\!\! &0,\!\!\!\!\!\!\!\! &0,\!\!\!\!\!\!\!\! &0,\!\!\!\!\!\!\!\! &0,\!\!\!\!\!\!\!\! &0,\!\!\!\!\!\!\!\! &0,\!\!\!\!\!\!\!\! &0,\!\!\!\!\!\!\!\! &0,\!\!\!\!\!\!\!\! &0,\!\!\!\!\!\!\!\! &0,\!\!\!\!\!\!\!\! &0,\!\!\!\!\!\!\!\! &0,\!\!\!\!\!\!\!\! &0,\!\!\!\!\!\!\!\! &0,\!\!\!\!\!\!\!\! &0,\!\!\!\!\!\!\!\! &0 \\
0,\!\!\!\!\!\!\!\! &0,\!\!\!\!\!\!\!\! &0,\!\!\!\!\!\!\!\! &0,\!\!\!\!\!\!\!\! &0,\!\!\!\!\!\!\!\! &0,\!\!\!\!\!\!\!\! &0,\!\!\!\!\!\!\!\! &0,\!\!\!\!\!\!\!\! &0,\!\!\!\!\!\!\!\! &0,\!\!\!\!\!\!\!\! &0,\!\!\!\!\!\!\!\! &0,\!\!\!\!\!\!\!\! &0,\!\!\!\!\!\!\!\! &0,\!\!\!\!\!\!\!\! &0,\!\!\!\!\!\!\!\! &0,\!\!\!\!\!\!\!\! &0,\!\!\!\!\!\!\!\! &0,\!\!\!\!\!\!\!\! &0,\!\!\!\!\!\!\!\! &0,\!\!\!\!\!\!\!\! &0,\!\!\!\!\!\!\!\! &0,\!\!\!\!\!\!\!\! &0,\!\!\!\!\!\!\!\! &0,\!\!\!\!\!\!\!\! &0,\!\!\!\!\!\!\!\! &0,\!\!\!\!\!\!\!\! &0 \\
\end{array}
\rp
\feqn
\eqn
c_{62}=\lp
\begin{array}{ccccccccccccccccccccccccccc}
0,\!\!\!\!\!\!\!\! &0,\!\!\!\!\!\!\!\! &0,\!\!\!\!\!\!\!\! &0,\!\!\!\!\!\!\!\! &0,\!\!\!\!\!\!\!\! &0,\!\!\!\!\!\!\!\! &0,\!\!\!\!\!\!\!\! &0,\!\!\!\!\!\!\!\! &0,\!\!\!\!\!\!\!\! &-\frac i{\sqrt 2},\!\!\!\!\!\!\!\! &0,\!\!\!\!\!\!\!\! &0,\!\!\!\!\!\!\!\! &0,\!\!\!\!\!\!\!\! &0,\!\!\!\!\!\!\!\! &0,\!\!\!\!\!\!\!\! &0,\!\!\!\!\!\!\!\! &0,\!\!\!\!\!\!\!\! &0,\!\!\!\!\!\!\!\! &0,\!\!\!\!\!\!\!\! &0,\!\!\!\!\!\!\!\! &0,\!\!\!\!\!\!\!\! &0,\!\!\!\!\!\!\!\! &0,\!\!\!\!\!\!\!\! &0,\!\!\!\!\!\!\!\! &0,\!\!\!\!\!\!\!\! &0,\!\!\!\!\!\!\!\! &0 \\
0,\!\!\!\!\!\!\!\! &0,\!\!\!\!\!\!\!\! &0,\!\!\!\!\!\!\!\! &0,\!\!\!\!\!\!\!\! &0,\!\!\!\!\!\!\!\! &0,\!\!\!\!\!\!\!\! &0,\!\!\!\!\!\!\!\! &0,\!\!\!\!\!\!\!\! &0,\!\!\!\!\!\!\!\! &0,\!\!\!\!\!\!\!\! &0,\!\!\!\!\!\!\!\! &0,\!\!\!\!\!\!\!\! &0,\!\!\!\!\!\!\!\! &0,\!\!\!\!\!\!\!\! &0,\!\!\!\!\!\!\!\! &0,\!\!\!\!\!\!\!\! &0,\!\!\!\!\!\!\!\! &0,\!\!\!\!\!\!\!\! &-\iem,\!\!\!\!\!\!\!\! &0,\!\!\!\!\!\!\!\! &0,\!\!\!\!\!\!\!\! &0,\!\!\!\!\!\!\!\! &0,\!\!\!\!\!\!\!\! &0,\!\!\!\!\!\!\!\! &0,\!\!\!\!\!\!\!\! &0,\!\!\!\!\!\!\!\! &0 \\
0,\!\!\!\!\!\!\!\! &0,\!\!\!\!\!\!\!\! &0,\!\!\!\!\!\!\!\! &0,\!\!\!\!\!\!\!\! &0,\!\!\!\!\!\!\!\! &0,\!\!\!\!\!\!\!\! &0,\!\!\!\!\!\!\!\! &0,\!\!\!\!\!\!\!\! &0,\!\!\!\!\!\!\!\! &0,\!\!\!\!\!\!\!\! &0,\!\!\!\!\!\!\!\! &0,\!\!\!\!\!\!\!\! &0,\!\!\!\!\!\!\!\! &0,\!\!\!\!\!\!\!\! &0,\!\!\!\!\!\!\!\! &0,\!\!\!\!\!\!\!\! &0,\!\!\!\!\!\!\!\! &0,\!\!\!\!\!\!\!\! &0,\!\!\!\!\!\!\!\! &\iem,\!\!\!\!\!\!\!\! &0,\!\!\!\!\!\!\!\! &0,\!\!\!\!\!\!\!\! &0,\!\!\!\!\!\!\!\! &0,\!\!\!\!\!\!\!\! &0,\!\!\!\!\!\!\!\! &0,\!\!\!\!\!\!\!\! &0 \\
\end{array}
\rp
\feqn
\eqn
c_{63}=\lp
\begin{array}{ccccccccccccccccccccccccccc}
0,\!\!\!\!\!\!\!\! &0,\!\!\!\!\!\!\!\! &0,\!\!\!\!\!\!\!\! &0,\!\!\!\!\!\!\!\! &0,\!\!\!\!\!\!\!\! &0,\!\!\!\!\!\!\!\! &0,\!\!\!\!\!\!\!\! &0,\!\!\!\!\!\!\!\! &0,\!\!\!\!\!\!\!\! &0,\!\!\!\!\!\!\!\! &-\frac i{\sqrt 2},\!\!\!\!\!\!\!\! &0,\!\!\!\!\!\!\!\! &0,\!\!\!\!\!\!\!\! &0,\!\!\!\!\!\!\!\! &0,\!\!\!\!\!\!\!\! &0,\!\!\!\!\!\!\!\! &0,\!\!\!\!\!\!\!\! &0,\!\!\!\!\!\!\!\! &0,\!\!\!\!\!\!\!\! &0,\!\!\!\!\!\!\!\! &0,\!\!\!\!\!\!\!\! &0,\!\!\!\!\!\!\!\! &0,\!\!\!\!\!\!\!\! &0,\!\!\!\!\!\!\!\! &0,\!\!\!\!\!\!\!\! &0,\!\!\!\!\!\!\!\! &0 \\
0,\!\!\!\!\!\!\!\! &0,\!\!\!\!\!\!\!\! &0,\!\!\!\!\!\!\!\! &0,\!\!\!\!\!\!\!\! &0,\!\!\!\!\!\!\!\! &0,\!\!\!\!\!\!\!\! &0,\!\!\!\!\!\!\!\! &0,\!\!\!\!\!\!\!\! &0,\!\!\!\!\!\!\!\! &0,\!\!\!\!\!\!\!\! &0,\!\!\!\!\!\!\!\! &0,\!\!\!\!\!\!\!\! &0,\!\!\!\!\!\!\!\! &0,\!\!\!\!\!\!\!\! &0,\!\!\!\!\!\!\!\! &0,\!\!\!\!\!\!\!\! &0,\!\!\!\!\!\!\!\! &0,\!\!\!\!\!\!\!\! &0,\!\!\!\!\!\!\!\! &-\iem,\!\!\!\!\!\!\!\! &0,\!\!\!\!\!\!\!\! &0,\!\!\!\!\!\!\!\! &0,\!\!\!\!\!\!\!\! &0,\!\!\!\!\!\!\!\! &0,\!\!\!\!\!\!\!\! &0,\!\!\!\!\!\!\!\! &0 \\
0,\!\!\!\!\!\!\!\! &0,\!\!\!\!\!\!\!\! &0,\!\!\!\!\!\!\!\! &0,\!\!\!\!\!\!\!\! &0,\!\!\!\!\!\!\!\! &0,\!\!\!\!\!\!\!\! &0,\!\!\!\!\!\!\!\! &0,\!\!\!\!\!\!\!\! &0,\!\!\!\!\!\!\!\! &0,\!\!\!\!\!\!\!\! &0,\!\!\!\!\!\!\!\! &0,\!\!\!\!\!\!\!\! &0,\!\!\!\!\!\!\!\! &0,\!\!\!\!\!\!\!\! &0,\!\!\!\!\!\!\!\! &0,\!\!\!\!\!\!\!\! &0,\!\!\!\!\!\!\!\! &0,\!\!\!\!\!\!\!\! &-\iem,\!\!\!\!\!\!\!\! &0,\!\!\!\!\!\!\!\! &0,\!\!\!\!\!\!\!\! &0,\!\!\!\!\!\!\!\! &0,\!\!\!\!\!\!\!\! &0,\!\!\!\!\!\!\!\! &0,\!\!\!\!\!\!\!\! &0,\!\!\!\!\!\!\!\! &0 \\
\end{array}
\rp
\feqn
\eqn
c_{64}=\lp
\begin{array}{ccccccccccccccccccccccccccc}
0,\!\!\!\!\!\!\!\! &0,\!\!\!\!\!\!\!\! &0,\!\!\!\!\!\!\!\! &0,\!\!\!\!\!\!\!\! &0,\!\!\!\!\!\!\!\! &0,\!\!\!\!\!\!\!\! &0,\!\!\!\!\!\!\!\! &0,\!\!\!\!\!\!\!\! &0,\!\!\!\!\!\!\!\! &0,\!\!\!\!\!\!\!\! &0,\!\!\!\!\!\!\!\! &-\frac i{\sqrt 2},\!\!\!\!\!\!\!\! &0,\!\!\!\!\!\!\!\! &0,\!\!\!\!\!\!\!\! &0,\!\!\!\!\!\!\!\! &0,\!\!\!\!\!\!\!\! &0,\!\!\!\!\!\!\!\! &0,\!\!\!\!\!\!\!\! &0,\!\!\!\!\!\!\!\! &0,\!\!\!\!\!\!\!\! &0,\!\!\!\!\!\!\!\! &0,\!\!\!\!\!\!\!\! &0,\!\!\!\!\!\!\!\! &0,\!\!\!\!\!\!\!\! &0,\!\!\!\!\!\!\!\! &0,\!\!\!\!\!\!\!\! &0 \\
0,\!\!\!\!\!\!\!\! &0,\!\!\!\!\!\!\!\! &0,\!\!\!\!\!\!\!\! &0,\!\!\!\!\!\!\!\! &0,\!\!\!\!\!\!\!\! &0,\!\!\!\!\!\!\!\! &0,\!\!\!\!\!\!\!\! &0,\!\!\!\!\!\!\!\! &0,\!\!\!\!\!\!\!\! &0,\!\!\!\!\!\!\!\! &0,\!\!\!\!\!\!\!\! &0,\!\!\!\!\!\!\!\! &0,\!\!\!\!\!\!\!\! &0,\!\!\!\!\!\!\!\! &0,\!\!\!\!\!\!\!\! &0,\!\!\!\!\!\!\!\! &0,\!\!\!\!\!\!\!\! &0,\!\!\!\!\!\!\!\! &0,\!\!\!\!\!\!\!\! &0,\!\!\!\!\!\!\!\! &-\iem,\!\!\!\!\!\!\!\! &0,\!\!\!\!\!\!\!\! &0,\!\!\!\!\!\!\!\! &0,\!\!\!\!\!\!\!\! &0,\!\!\!\!\!\!\!\! &0,\!\!\!\!\!\!\!\! &0 \\
0,\!\!\!\!\!\!\!\! &0,\!\!\!\!\!\!\!\! &0,\!\!\!\!\!\!\!\! &0,\!\!\!\!\!\!\!\! &0,\!\!\!\!\!\!\!\! &0,\!\!\!\!\!\!\!\! &0,\!\!\!\!\!\!\!\! &0,\!\!\!\!\!\!\!\! &0,\!\!\!\!\!\!\!\! &0,\!\!\!\!\!\!\!\! &0,\!\!\!\!\!\!\!\! &0,\!\!\!\!\!\!\!\! &0,\!\!\!\!\!\!\!\! &0,\!\!\!\!\!\!\!\! &0,\!\!\!\!\!\!\!\! &0,\!\!\!\!\!\!\!\! &0,\!\!\!\!\!\!\!\! &0,\!\!\!\!\!\!\!\! &0,\!\!\!\!\!\!\!\! &0,\!\!\!\!\!\!\!\! &0,\!\!\!\!\!\!\!\! &0,\!\!\!\!\!\!\!\! &\iem,\!\!\!\!\!\!\!\! &0,\!\!\!\!\!\!\!\! &0,\!\!\!\!\!\!\!\! &0,\!\!\!\!\!\!\!\! &0 \\
\end{array}
\rp
\feqn
\eqn
c_{65}=\lp
\begin{array}{ccccccccccccccccccccccccccc}
0,\!\!\!\!\!\!\!\! &0,\!\!\!\!\!\!\!\! &0,\!\!\!\!\!\!\!\! &0,\!\!\!\!\!\!\!\! &0,\!\!\!\!\!\!\!\! &0,\!\!\!\!\!\!\!\! &0,\!\!\!\!\!\!\!\! &0,\!\!\!\!\!\!\!\! &0,\!\!\!\!\!\!\!\! &0,\!\!\!\!\!\!\!\! &0,\!\!\!\!\!\!\!\! &0,\!\!\!\!\!\!\!\! &-i\frac 1{\sqrt 2},\!\!\!\!\!\!\!\! &0,\!\!\!\!\!\!\!\! &0,\!\!\!\!\!\!\!\! &0,\!\!\!\!\!\!\!\! &0,\!\!\!\!\!\!\!\! &0,\!\!\!\!\!\!\!\! &0,\!\!\!\!\!\!\!\! &0,\!\!\!\!\!\!\!\! &0,\!\!\!\!\!\!\!\! &0,\!\!\!\!\!\!\!\! &0,\!\!\!\!\!\!\!\! &0,\!\!\!\!\!\!\!\! &0,\!\!\!\!\!\!\!\! &0,\!\!\!\!\!\!\!\! &0 \\
0,\!\!\!\!\!\!\!\! &0,\!\!\!\!\!\!\!\! &0,\!\!\!\!\!\!\!\! &0,\!\!\!\!\!\!\!\! &0,\!\!\!\!\!\!\!\! &0,\!\!\!\!\!\!\!\! &0,\!\!\!\!\!\!\!\! &0,\!\!\!\!\!\!\!\! &0,\!\!\!\!\!\!\!\! &0,\!\!\!\!\!\!\!\! &0,\!\!\!\!\!\!\!\! &0,\!\!\!\!\!\!\!\! &0,\!\!\!\!\!\!\!\! &0,\!\!\!\!\!\!\!\! &0,\!\!\!\!\!\!\!\! &0,\!\!\!\!\!\!\!\! &0,\!\!\!\!\!\!\!\! &0,\!\!\!\!\!\!\!\! &0,\!\!\!\!\!\!\!\! &0,\!\!\!\!\!\!\!\! &0,\!\!\!\!\!\!\!\! &-\iem,\!\!\!\!\!\!\!\! &0,\!\!\!\!\!\!\!\! &0,\!\!\!\!\!\!\!\! &0,\!\!\!\!\!\!\!\! &0,\!\!\!\!\!\!\!\! &0 \\
0,\!\!\!\!\!\!\!\! &0,\!\!\!\!\!\!\!\! &0,\!\!\!\!\!\!\!\! &0,\!\!\!\!\!\!\!\! &0,\!\!\!\!\!\!\!\! &0,\!\!\!\!\!\!\!\! &0,\!\!\!\!\!\!\!\! &0,\!\!\!\!\!\!\!\! &0,\!\!\!\!\!\!\!\! &0,\!\!\!\!\!\!\!\! &0,\!\!\!\!\!\!\!\! &0,\!\!\!\!\!\!\!\! &0,\!\!\!\!\!\!\!\! &0,\!\!\!\!\!\!\!\! &0,\!\!\!\!\!\!\!\! &0,\!\!\!\!\!\!\!\! &0,\!\!\!\!\!\!\!\! &0,\!\!\!\!\!\!\!\! &0,\!\!\!\!\!\!\!\! &0,\!\!\!\!\!\!\!\! &0,\!\!\!\!\!\!\!\! &0,\!\!\!\!\!\!\!\! &0,\!\!\!\!\!\!\!\! &0,\!\!\!\!\!\!\!\! &0,\!\!\!\!\!\!\!\! &\iem,\!\!\!\!\!\!\!\! &0 \\
\end{array}
\rp
\feqn
\eqn
c_{66}=\lp
\begin{array}{ccccccccccccccccccccccccccc}
0,\!\!\!\!\!\!\!\! &0,\!\!\!\!\!\!\!\! &0,\!\!\!\!\!\!\!\! &0,\!\!\!\!\!\!\!\! &0,\!\!\!\!\!\!\!\! &0,\!\!\!\!\!\!\!\! &0,\!\!\!\!\!\!\!\! &0,\!\!\!\!\!\!\!\! &0,\!\!\!\!\!\!\!\! &0,\!\!\!\!\!\!\!\! &0,\!\!\!\!\!\!\!\! &0,\!\!\!\!\!\!\!\! &0,\!\!\!\!\!\!\!\! &-\frac i{\sqrt 2},\!\!\!\!\!\!\!\! &0,\!\!\!\!\!\!\!\! &0,\!\!\!\!\!\!\!\! &0,\!\!\!\!\!\!\!\! &0,\!\!\!\!\!\!\!\! &0,\!\!\!\!\!\!\!\! &0,\!\!\!\!\!\!\!\! &0,\!\!\!\!\!\!\!\! &0,\!\!\!\!\!\!\!\! &0,\!\!\!\!\!\!\!\! &0,\!\!\!\!\!\!\!\! &0,\!\!\!\!\!\!\!\! &0,\!\!\!\!\!\!\!\! &0 \\
0,\!\!\!\!\!\!\!\! &0,\!\!\!\!\!\!\!\! &0,\!\!\!\!\!\!\!\! &0,\!\!\!\!\!\!\!\! &0,\!\!\!\!\!\!\!\! &0,\!\!\!\!\!\!\!\! &0,\!\!\!\!\!\!\!\! &0,\!\!\!\!\!\!\!\! &0,\!\!\!\!\!\!\!\! &0,\!\!\!\!\!\!\!\! &0,\!\!\!\!\!\!\!\! &0,\!\!\!\!\!\!\!\! &0,\!\!\!\!\!\!\!\! &0,\!\!\!\!\!\!\!\! &0,\!\!\!\!\!\!\!\! &0,\!\!\!\!\!\!\!\! &0,\!\!\!\!\!\!\!\! &0,\!\!\!\!\!\!\!\! &0,\!\!\!\!\!\!\!\! &0,\!\!\!\!\!\!\!\! &0,\!\!\!\!\!\!\!\! &0,\!\!\!\!\!\!\!\! &-\iem,\!\!\!\!\!\!\!\! &0,\!\!\!\!\!\!\!\! &0,\!\!\!\!\!\!\!\! &0,\!\!\!\!\!\!\!\! &0 \\
0,\!\!\!\!\!\!\!\! &0,\!\!\!\!\!\!\!\! &0,\!\!\!\!\!\!\!\! &0,\!\!\!\!\!\!\!\! &0,\!\!\!\!\!\!\!\! &0,\!\!\!\!\!\!\!\! &0,\!\!\!\!\!\!\!\! &0,\!\!\!\!\!\!\!\! &0,\!\!\!\!\!\!\!\! &0,\!\!\!\!\!\!\!\! &0,\!\!\!\!\!\!\!\! &0,\!\!\!\!\!\!\!\! &0,\!\!\!\!\!\!\!\! &0,\!\!\!\!\!\!\!\! &0,\!\!\!\!\!\!\!\! &0,\!\!\!\!\!\!\!\! &0,\!\!\!\!\!\!\!\! &0,\!\!\!\!\!\!\!\! &0,\!\!\!\!\!\!\!\! &0,\!\!\!\!\!\!\!\! &-\iem,\!\!\!\!\!\!\!\! &0,\!\!\!\!\!\!\!\! &0,\!\!\!\!\!\!\!\! &0,\!\!\!\!\!\!\!\! &0,\!\!\!\!\!\!\!\! &0,\!\!\!\!\!\!\!\! &0 \\
\end{array}
\rp
\feqn
\eqn
c_{67}=\lp
\begin{array}{ccccccccccccccccccccccccccc}
0,\!\!\!\!\!\!\!\! &0,\!\!\!\!\!\!\!\! &0,\!\!\!\!\!\!\!\! &0,\!\!\!\!\!\!\!\! &0,\!\!\!\!\!\!\!\! &0,\!\!\!\!\!\!\!\! &0,\!\!\!\!\!\!\!\! &0,\!\!\!\!\!\!\!\! &0,\!\!\!\!\!\!\!\! &0,\!\!\!\!\!\!\!\! &0,\!\!\!\!\!\!\!\! &0,\!\!\!\!\!\!\!\! &0,\!\!\!\!\!\!\!\! &0,\!\!\!\!\!\!\!\! &-\frac i{\sqrt 2},\!\!\!\!\!\!\!\! &0,\!\!\!\!\!\!\!\! &0,\!\!\!\!\!\!\!\! &0,\!\!\!\!\!\!\!\! &0,\!\!\!\!\!\!\!\! &0,\!\!\!\!\!\!\!\! &0,\!\!\!\!\!\!\!\! &0,\!\!\!\!\!\!\!\! &0,\!\!\!\!\!\!\!\! &0,\!\!\!\!\!\!\!\! &0,\!\!\!\!\!\!\!\! &0,\!\!\!\!\!\!\!\! &0 \\
0,\!\!\!\!\!\!\!\! &0,\!\!\!\!\!\!\!\! &0,\!\!\!\!\!\!\!\! &0,\!\!\!\!\!\!\!\! &0,\!\!\!\!\!\!\!\! &0,\!\!\!\!\!\!\!\! &0,\!\!\!\!\!\!\!\! &0,\!\!\!\!\!\!\!\! &0,\!\!\!\!\!\!\!\! &0,\!\!\!\!\!\!\!\! &0,\!\!\!\!\!\!\!\! &0,\!\!\!\!\!\!\!\! &0,\!\!\!\!\!\!\!\! &0,\!\!\!\!\!\!\!\! &0,\!\!\!\!\!\!\!\! &0,\!\!\!\!\!\!\!\! &0,\!\!\!\!\!\!\!\! &0,\!\!\!\!\!\!\!\! &0,\!\!\!\!\!\!\!\! &0,\!\!\!\!\!\!\!\! &0,\!\!\!\!\!\!\!\! &0,\!\!\!\!\!\!\!\! &0,\!\!\!\!\!\!\!\! &-\iem,\!\!\!\!\!\!\!\! &0,\!\!\!\!\!\!\!\! &0,\!\!\!\!\!\!\!\! &0 \\
0,\!\!\!\!\!\!\!\! &0,\!\!\!\!\!\!\!\! &0,\!\!\!\!\!\!\!\! &0,\!\!\!\!\!\!\!\! &0,\!\!\!\!\!\!\!\! &0,\!\!\!\!\!\!\!\! &0,\!\!\!\!\!\!\!\! &0,\!\!\!\!\!\!\!\! &0,\!\!\!\!\!\!\!\! &0,\!\!\!\!\!\!\!\! &0,\!\!\!\!\!\!\!\! &0,\!\!\!\!\!\!\!\! &0,\!\!\!\!\!\!\!\! &0,\!\!\!\!\!\!\!\! &0,\!\!\!\!\!\!\!\! &0,\!\!\!\!\!\!\!\! &0,\!\!\!\!\!\!\!\! &0,\!\!\!\!\!\!\!\! &0,\!\!\!\!\!\!\!\! &0,\!\!\!\!\!\!\!\! &0,\!\!\!\!\!\!\!\! &0,\!\!\!\!\!\!\!\! &0,\!\!\!\!\!\!\!\! &0,\!\!\!\!\!\!\!\! &\iem,\!\!\!\!\!\!\!\! &0,\!\!\!\!\!\!\!\! &0 \\
\end{array}
\rp
\feqn
\eqn
c_{68}=\lp
\begin{array}{ccccccccccccccccccccccccccc}
0,\!\!\!\!\!\!\!\! &0,\!\!\!\!\!\!\!\! &0,\!\!\!\!\!\!\!\! &0,\!\!\!\!\!\!\!\! &0,\!\!\!\!\!\!\!\! &0,\!\!\!\!\!\!\!\! &0,\!\!\!\!\!\!\!\! &0,\!\!\!\!\!\!\!\! &0,\!\!\!\!\!\!\!\! &0,\!\!\!\!\!\!\!\! &0,\!\!\!\!\!\!\!\! &0,\!\!\!\!\!\!\!\! &0,\!\!\!\!\!\!\!\! &0,\!\!\!\!\!\!\!\! &0,\!\!\!\!\!\!\!\! &-\frac i{\sqrt 2},\!\!\!\!\!\!\!\! &0,\!\!\!\!\!\!\!\! &0,\!\!\!\!\!\!\!\! &0,\!\!\!\!\!\!\!\! &0,\!\!\!\!\!\!\!\! &0,\!\!\!\!\!\!\!\! &0,\!\!\!\!\!\!\!\! &0,\!\!\!\!\!\!\!\! &0,\!\!\!\!\!\!\!\! &0,\!\!\!\!\!\!\!\! &0,\!\!\!\!\!\!\!\! &0 \\
0,\!\!\!\!\!\!\!\! &0,\!\!\!\!\!\!\!\! &0,\!\!\!\!\!\!\!\! &0,\!\!\!\!\!\!\!\! &0,\!\!\!\!\!\!\!\! &0,\!\!\!\!\!\!\!\! &0,\!\!\!\!\!\!\!\! &0,\!\!\!\!\!\!\!\! &0,\!\!\!\!\!\!\!\! &0,\!\!\!\!\!\!\!\! &0,\!\!\!\!\!\!\!\! &0,\!\!\!\!\!\!\!\! &0,\!\!\!\!\!\!\!\! &0,\!\!\!\!\!\!\!\! &0,\!\!\!\!\!\!\!\! &0,\!\!\!\!\!\!\!\! &0,\!\!\!\!\!\!\!\! &0,\!\!\!\!\!\!\!\! &0,\!\!\!\!\!\!\!\! &0,\!\!\!\!\!\!\!\! &0,\!\!\!\!\!\!\!\! &0,\!\!\!\!\!\!\!\! &0,\!\!\!\!\!\!\!\! &0,\!\!\!\!\!\!\!\! &-\iem,\!\!\!\!\!\!\!\! &0,\!\!\!\!\!\!\!\! &0 \\
0,\!\!\!\!\!\!\!\! &0,\!\!\!\!\!\!\!\! &0,\!\!\!\!\!\!\!\! &0,\!\!\!\!\!\!\!\! &0,\!\!\!\!\!\!\!\! &0,\!\!\!\!\!\!\!\! &0,\!\!\!\!\!\!\!\! &0,\!\!\!\!\!\!\!\! &0,\!\!\!\!\!\!\!\! &0,\!\!\!\!\!\!\!\! &0,\!\!\!\!\!\!\!\! &0,\!\!\!\!\!\!\!\! &0,\!\!\!\!\!\!\!\! &0,\!\!\!\!\!\!\!\! &0,\!\!\!\!\!\!\!\! &0,\!\!\!\!\!\!\!\! &0,\!\!\!\!\!\!\!\! &0,\!\!\!\!\!\!\!\! &0,\!\!\!\!\!\!\!\! &0,\!\!\!\!\!\!\!\! &0,\!\!\!\!\!\!\!\! &0,\!\!\!\!\!\!\!\! &0,\!\!\!\!\!\!\!\! &-\iem,\!\!\!\!\!\!\!\! &0,\!\!\!\!\!\!\!\! &0,\!\!\!\!\!\!\!\! &0 \\
\end{array}
\rp
\feqn
\eqn
c_{69}=\lp
\begin{array}{ccccccccccccccccccccccccccc}
0,\!\!\!\!\!\!\!\! &0,\!\!\!\!\!\!\!\! &0,\!\!\!\!\!\!\!\! &0,\!\!\!\!\!\!\!\! &0,\!\!\!\!\!\!\!\! &0,\!\!\!\!\!\!\!\! &0,\!\!\!\!\!\!\!\! &0,\!\!\!\!\!\!\!\! &0,\!\!\!\!\!\!\!\! &0,\!\!\!\!\!\!\!\! &0,\!\!\!\!\!\!\!\! &0,\!\!\!\!\!\!\!\! &0,\!\!\!\!\!\!\!\! &0,\!\!\!\!\!\!\!\! &0,\!\!\!\!\!\!\!\! &0,\!\!\!\!\!\!\!\! &-\frac i{\sqrt 2},\!\!\!\!\!\!\!\! &0,\!\!\!\!\!\!\!\! &0,\!\!\!\!\!\!\!\! &0,\!\!\!\!\!\!\!\! &0,\!\!\!\!\!\!\!\! &0,\!\!\!\!\!\!\!\! &0,\!\!\!\!\!\!\!\! &0,\!\!\!\!\!\!\!\! &0,\!\!\!\!\!\!\!\! &0,\!\!\!\!\!\!\!\! &0 \\
0,\!\!\!\!\!\!\!\! &0,\!\!\!\!\!\!\!\! &0,\!\!\!\!\!\!\!\! &0,\!\!\!\!\!\!\!\! &0,\!\!\!\!\!\!\!\! &0,\!\!\!\!\!\!\!\! &0,\!\!\!\!\!\!\!\! &0,\!\!\!\!\!\!\!\! &0,\!\!\!\!\!\!\!\! &0,\!\!\!\!\!\!\!\! &0,\!\!\!\!\!\!\!\! &0,\!\!\!\!\!\!\!\! &0,\!\!\!\!\!\!\!\! &0,\!\!\!\!\!\!\!\! &0,\!\!\!\!\!\!\!\! &0,\!\!\!\!\!\!\!\! &0,\!\!\!\!\!\!\!\! &0,\!\!\!\!\!\!\!\! &0,\!\!\!\!\!\!\!\! &0,\!\!\!\!\!\!\!\! &0,\!\!\!\!\!\!\!\! &0,\!\!\!\!\!\!\!\! &0,\!\!\!\!\!\!\!\! &0,\!\!\!\!\!\!\!\! &0,\!\!\!\!\!\!\!\! &-\iem,\!\!\!\!\!\!\!\! &0 \\
0,\!\!\!\!\!\!\!\! &0,\!\!\!\!\!\!\!\! &0,\!\!\!\!\!\!\!\! &0,\!\!\!\!\!\!\!\! &0,\!\!\!\!\!\!\!\! &0,\!\!\!\!\!\!\!\! &0,\!\!\!\!\!\!\!\! &0,\!\!\!\!\!\!\!\! &0,\!\!\!\!\!\!\!\! &0,\!\!\!\!\!\!\!\! &0,\!\!\!\!\!\!\!\! &0,\!\!\!\!\!\!\!\! &0,\!\!\!\!\!\!\!\! &0,\!\!\!\!\!\!\!\! &0,\!\!\!\!\!\!\!\! &0,\!\!\!\!\!\!\!\! &0,\!\!\!\!\!\!\!\! &0,\!\!\!\!\!\!\!\! &0,\!\!\!\!\!\!\!\! &0,\!\!\!\!\!\!\!\! &0,\!\!\!\!\!\!\!\! &-\iem,\!\!\!\!\!\!\!\! &0,\!\!\!\!\!\!\!\! &0,\!\!\!\!\!\!\!\! &0,\!\!\!\!\!\!\!\! &0,\!\!\!\!\!\!\!\! &0 \\
\end{array}
\rp
\feqn
\eqn
c_{70}=\lp
\begin{array}{ccccccccccccccccccccccccccc}
-\frac \kappa{2},\!\!\!\!\!\!\!\! &0,\!\!\!\!\!\!\!\! &0,\!\!\!\!\!\!\!\! &0,\!\!\!\!\!\!\!\! &0,\!\!\!\!\!\!\!\! &0,\!\!\!\!\!\!\!\! &0,\!\!\!\!\!\!\!\! &0,\!\!\!\!\!\!\!\! &0,\!\!\!\!\!\!\!\! &0,\!\!\!\!\!\!\!\! &0,\!\!\!\!\!\!\!\! &0,\!\!\!\!\!\!\!\! &0,\!\!\!\!\!\!\!\! &0,\!\!\!\!\!\!\!\! &0,\!\!\!\!\!\!\!\! &0,\!\!\!\!\!\!\!\! &0,\!\!\!\!\!\!\!\! &\frac i2,\!\!\!\!\!\!\!\! &0,\!\!\!\!\!\!\!\! &0,\!\!\!\!\!\!\!\! &0,\!\!\!\!\!\!\!\! &0,\!\!\!\!\!\!\!\! &0,\!\!\!\!\!\!\!\! &0,\!\!\!\!\!\!\!\! &0,\!\!\!\!\!\!\!\! &0,\!\!\!\!\!\!\!\! &\frac i{\sqrt 2} \\
0,\!\!\!\!\!\!\!\! &-\frac \kappa{2},\!\!\!\!\!\!\!\! &0,\!\!\!\!\!\!\!\! &0,\!\!\!\!\!\!\!\! &0,\!\!\!\!\!\!\!\! &0,\!\!\!\!\!\!\!\! &0,\!\!\!\!\!\!\!\! &0,\!\!\!\!\!\!\!\! &0,\!\!\!\!\!\!\!\! &0,\!\!\!\!\!\!\!\! &0,\!\!\!\!\!\!\!\! &0,\!\!\!\!\!\!\!\! &0,\!\!\!\!\!\!\!\! &0,\!\!\!\!\!\!\!\! &0,\!\!\!\!\!\!\!\! &0,\!\!\!\!\!\!\!\! &0,\!\!\!\!\!\!\!\! &0,\!\!\!\!\!\!\!\! &0,\!\!\!\!\!\!\!\! &0,\!\!\!\!\!\!\!\! &0,\!\!\!\!\!\!\!\! &0,\!\!\!\!\!\!\!\! &0,\!\!\!\!\!\!\!\! &0,\!\!\!\!\!\!\!\! &0,\!\!\!\!\!\!\!\! &0,\!\!\!\!\!\!\!\! &0 \\
0,\!\!\!\!\!\!\!\! &0,\!\!\!\!\!\!\!\! &-\frac \kappa{2},\!\!\!\!\!\!\!\! &0,\!\!\!\!\!\!\!\! &0,\!\!\!\!\!\!\!\! &0,\!\!\!\!\!\!\!\! &0,\!\!\!\!\!\!\!\! &0,\!\!\!\!\!\!\!\! &0,\!\!\!\!\!\!\!\! &0,\!\!\!\!\!\!\!\! &0,\!\!\!\!\!\!\!\! &0,\!\!\!\!\!\!\!\! &0,\!\!\!\!\!\!\!\! &0,\!\!\!\!\!\!\!\! &0,\!\!\!\!\!\!\!\! &0,\!\!\!\!\!\!\!\! &0,\!\!\!\!\!\!\!\! &0,\!\!\!\!\!\!\!\! &0,\!\!\!\!\!\!\!\! &0,\!\!\!\!\!\!\!\! &0,\!\!\!\!\!\!\!\! &0,\!\!\!\!\!\!\!\! &0,\!\!\!\!\!\!\!\! &0,\!\!\!\!\!\!\!\! &0,\!\!\!\!\!\!\!\! &0,\!\!\!\!\!\!\!\! &0 \\
\end{array}
\rp
\feqn
where $\kappa=\frac i{\sqrt 3}$\ ,
\eqn
c_{71}=\lp
\begin{array}{ccccccccccccccccccccccccccc}
0,\!\!\!\!\!\!\!\! &0,\!\!\!\!\!\!\!\! &0,\!\!\!\!\!\!\!\! &0,\!\!\!\!\!\!\!\! &0,\!\!\!\!\!\!\!\! &0,\!\!\!\!\!\!\!\! &0,\!\!\!\!\!\!\!\! &0,\!\!\!\!\!\!\!\! &0,\!\!\!\!\!\!\!\! &0,\!\!\!\!\!\!\!\! &0,\!\!\!\!\!\!\!\! &0,\!\!\!\!\!\!\!\! &0,\!\!\!\!\!\!\!\! &0,\!\!\!\!\!\!\!\! &0,\!\!\!\!\!\!\!\! &0,\!\!\!\!\!\!\!\! &0,\!\!\!\!\!\!\!\! &0,\!\!\!\!\!\!\!\! &\frac i{\sqrt 2},\!\!\!\!\!\!\!\! &0,\!\!\!\!\!\!\!\! &0,\!\!\!\!\!\!\!\! &0,\!\!\!\!\!\!\!\! &0,\!\!\!\!\!\!\!\! &0,\!\!\!\!\!\!\!\! &0,\!\!\!\!\!\!\!\! &0,\!\!\!\!\!\!\!\! &0 \\
0,\!\!\!\!\!\!\!\! &0,\!\!\!\!\!\!\!\! &0,\!\!\!\!\!\!\!\! &0,\!\!\!\!\!\!\!\! &0,\!\!\!\!\!\!\!\! &0,\!\!\!\!\!\!\!\! &0,\!\!\!\!\!\!\!\! &0,\!\!\!\!\!\!\!\! &0,\!\!\!\!\!\!\!\! &-\iem,\!\!\!\!\!\!\!\! &0,\!\!\!\!\!\!\!\! &0,\!\!\!\!\!\!\!\! &0,\!\!\!\!\!\!\!\! &0,\!\!\!\!\!\!\!\! &0,\!\!\!\!\!\!\!\! &0,\!\!\!\!\!\!\!\! &0,\!\!\!\!\!\!\!\! &0,\!\!\!\!\!\!\!\! &0,\!\!\!\!\!\!\!\! &0,\!\!\!\!\!\!\!\! &0,\!\!\!\!\!\!\!\! &0,\!\!\!\!\!\!\!\! &0,\!\!\!\!\!\!\!\! &0,\!\!\!\!\!\!\!\! &0,\!\!\!\!\!\!\!\! &0,\!\!\!\!\!\!\!\! &0 \\
0,\!\!\!\!\!\!\!\! &0,\!\!\!\!\!\!\!\! &0,\!\!\!\!\!\!\!\! &0,\!\!\!\!\!\!\!\! &0,\!\!\!\!\!\!\!\! &0,\!\!\!\!\!\!\!\! &0,\!\!\!\!\!\!\!\! &0,\!\!\!\!\!\!\!\! &0,\!\!\!\!\!\!\!\! &0,\!\!\!\!\!\!\!\! &-\iem,\!\!\!\!\!\!\!\! &0,\!\!\!\!\!\!\!\! &0,\!\!\!\!\!\!\!\! &0,\!\!\!\!\!\!\!\! &0,\!\!\!\!\!\!\!\! &0,\!\!\!\!\!\!\!\! &0,\!\!\!\!\!\!\!\! &0,\!\!\!\!\!\!\!\! &0,\!\!\!\!\!\!\!\! &0,\!\!\!\!\!\!\!\! &0,\!\!\!\!\!\!\!\! &0,\!\!\!\!\!\!\!\! &0,\!\!\!\!\!\!\!\! &0,\!\!\!\!\!\!\!\! &0,\!\!\!\!\!\!\!\! &0,\!\!\!\!\!\!\!\! &0 \\
\end{array}
\rp
\feqn
\eqn
c_{72}=\lp
\begin{array}{ccccccccccccccccccccccccccc}
0,\!\!\!\!\!\!\!\! &0,\!\!\!\!\!\!\!\! &0,\!\!\!\!\!\!\!\! &0,\!\!\!\!\!\!\!\! &0,\!\!\!\!\!\!\!\! &0,\!\!\!\!\!\!\!\! &0,\!\!\!\!\!\!\!\! &0,\!\!\!\!\!\!\!\! &0,\!\!\!\!\!\!\!\! &0,\!\!\!\!\!\!\!\! &0,\!\!\!\!\!\!\!\! &0,\!\!\!\!\!\!\!\! &0,\!\!\!\!\!\!\!\! &0,\!\!\!\!\!\!\!\! &0,\!\!\!\!\!\!\!\! &0,\!\!\!\!\!\!\!\! &0,\!\!\!\!\!\!\!\! &0,\!\!\!\!\!\!\!\! &0,\!\!\!\!\!\!\!\! &\frac i{\sqrt 2},\!\!\!\!\!\!\!\! &0,\!\!\!\!\!\!\!\! &0,\!\!\!\!\!\!\!\! &0,\!\!\!\!\!\!\!\! &0,\!\!\!\!\!\!\!\! &0,\!\!\!\!\!\!\!\! &0,\!\!\!\!\!\!\!\! &0 \\
0,\!\!\!\!\!\!\!\! &0,\!\!\!\!\!\!\!\! &0,\!\!\!\!\!\!\!\! &0,\!\!\!\!\!\!\!\! &0,\!\!\!\!\!\!\!\! &0,\!\!\!\!\!\!\!\! &0,\!\!\!\!\!\!\!\! &0,\!\!\!\!\!\!\!\! &0,\!\!\!\!\!\!\!\! &0,\!\!\!\!\!\!\!\! &-\iem,\!\!\!\!\!\!\!\! &0,\!\!\!\!\!\!\!\! &0,\!\!\!\!\!\!\!\! &0,\!\!\!\!\!\!\!\! &0,\!\!\!\!\!\!\!\! &0,\!\!\!\!\!\!\!\! &0,\!\!\!\!\!\!\!\! &0,\!\!\!\!\!\!\!\! &0,\!\!\!\!\!\!\!\! &0,\!\!\!\!\!\!\!\! &0,\!\!\!\!\!\!\!\! &0,\!\!\!\!\!\!\!\! &0,\!\!\!\!\!\!\!\! &0,\!\!\!\!\!\!\!\! &0,\!\!\!\!\!\!\!\! &0,\!\!\!\!\!\!\!\! &0 \\
0,\!\!\!\!\!\!\!\! &0,\!\!\!\!\!\!\!\! &0,\!\!\!\!\!\!\!\! &0,\!\!\!\!\!\!\!\! &0,\!\!\!\!\!\!\!\! &0,\!\!\!\!\!\!\!\! &0,\!\!\!\!\!\!\!\! &0,\!\!\!\!\!\!\!\! &0,\!\!\!\!\!\!\!\! &\iem,\!\!\!\!\!\!\!\! &0,\!\!\!\!\!\!\!\! &0,\!\!\!\!\!\!\!\! &0,\!\!\!\!\!\!\!\! &0,\!\!\!\!\!\!\!\! &0,\!\!\!\!\!\!\!\! &0,\!\!\!\!\!\!\!\! &0,\!\!\!\!\!\!\!\! &0,\!\!\!\!\!\!\!\! &0,\!\!\!\!\!\!\!\! &0,\!\!\!\!\!\!\!\! &0,\!\!\!\!\!\!\!\! &0,\!\!\!\!\!\!\!\! &0,\!\!\!\!\!\!\!\! &0,\!\!\!\!\!\!\!\! &0,\!\!\!\!\!\!\!\! &0,\!\!\!\!\!\!\!\! &0 \\
\end{array}
\rp
\feqn
\eqn
c_{73}=\lp
\begin{array}{ccccccccccccccccccccccccccc}
0,\!\!\!\!\!\!\!\! &0,\!\!\!\!\!\!\!\! &0,\!\!\!\!\!\!\!\! &0,\!\!\!\!\!\!\!\! &0,\!\!\!\!\!\!\!\! &0,\!\!\!\!\!\!\!\! &0,\!\!\!\!\!\!\!\! &0,\!\!\!\!\!\!\!\! &0,\!\!\!\!\!\!\!\! &0,\!\!\!\!\!\!\!\! &0,\!\!\!\!\!\!\!\! &0,\!\!\!\!\!\!\!\! &0,\!\!\!\!\!\!\!\! &0,\!\!\!\!\!\!\!\! &0,\!\!\!\!\!\!\!\! &0,\!\!\!\!\!\!\!\! &0,\!\!\!\!\!\!\!\! &0,\!\!\!\!\!\!\!\! &0,\!\!\!\!\!\!\!\! &0,\!\!\!\!\!\!\!\! &\frac i{\sqrt 2},\!\!\!\!\!\!\!\! &0,\!\!\!\!\!\!\!\! &0,\!\!\!\!\!\!\!\! &0,\!\!\!\!\!\!\!\! &0,\!\!\!\!\!\!\!\! &0,\!\!\!\!\!\!\!\! &0 \\
0,\!\!\!\!\!\!\!\! &0,\!\!\!\!\!\!\!\! &0,\!\!\!\!\!\!\!\! &0,\!\!\!\!\!\!\!\! &0,\!\!\!\!\!\!\!\! &0,\!\!\!\!\!\!\!\! &0,\!\!\!\!\!\!\!\! &0,\!\!\!\!\!\!\!\! &0,\!\!\!\!\!\!\!\! &0,\!\!\!\!\!\!\!\! &0,\!\!\!\!\!\!\!\! &-\iem,\!\!\!\!\!\!\!\! &0,\!\!\!\!\!\!\!\! &0,\!\!\!\!\!\!\!\! &0,\!\!\!\!\!\!\!\! &0,\!\!\!\!\!\!\!\! &0,\!\!\!\!\!\!\!\! &0,\!\!\!\!\!\!\!\! &0,\!\!\!\!\!\!\!\! &0,\!\!\!\!\!\!\!\! &0,\!\!\!\!\!\!\!\! &0,\!\!\!\!\!\!\!\! &0,\!\!\!\!\!\!\!\! &0,\!\!\!\!\!\!\!\! &0,\!\!\!\!\!\!\!\! &0,\!\!\!\!\!\!\!\! &0 \\
0,\!\!\!\!\!\!\!\! &0,\!\!\!\!\!\!\!\! &0,\!\!\!\!\!\!\!\! &0,\!\!\!\!\!\!\!\! &0,\!\!\!\!\!\!\!\! &0,\!\!\!\!\!\!\!\! &0,\!\!\!\!\!\!\!\! &0,\!\!\!\!\!\!\!\! &0,\!\!\!\!\!\!\!\! &0,\!\!\!\!\!\!\!\! &0,\!\!\!\!\!\!\!\! &0,\!\!\!\!\!\!\!\! &0,\!\!\!\!\!\!\!\! &-\iem,\!\!\!\!\!\!\!\! &0,\!\!\!\!\!\!\!\! &0,\!\!\!\!\!\!\!\! &0,\!\!\!\!\!\!\!\! &0,\!\!\!\!\!\!\!\! &0,\!\!\!\!\!\!\!\! &0,\!\!\!\!\!\!\!\! &0,\!\!\!\!\!\!\!\! &0,\!\!\!\!\!\!\!\! &0,\!\!\!\!\!\!\!\! &0,\!\!\!\!\!\!\!\! &0,\!\!\!\!\!\!\!\! &0,\!\!\!\!\!\!\!\! &0 \\
\end{array}
\rp
\feqn
\eqn
c_{74}=\lp
\begin{array}{ccccccccccccccccccccccccccc}
0,\!\!\!\!\!\!\!\! &0,\!\!\!\!\!\!\!\! &0,\!\!\!\!\!\!\!\! &0,\!\!\!\!\!\!\!\! &0,\!\!\!\!\!\!\!\! &0,\!\!\!\!\!\!\!\! &0,\!\!\!\!\!\!\!\! &0,\!\!\!\!\!\!\!\! &0,\!\!\!\!\!\!\!\! &0,\!\!\!\!\!\!\!\! &0,\!\!\!\!\!\!\!\! &0,\!\!\!\!\!\!\!\! &0,\!\!\!\!\!\!\!\! &0,\!\!\!\!\!\!\!\! &0,\!\!\!\!\!\!\!\! &0,\!\!\!\!\!\!\!\! &0,\!\!\!\!\!\!\!\! &0,\!\!\!\!\!\!\!\! &0,\!\!\!\!\!\!\!\! &0,\!\!\!\!\!\!\!\! &0,\!\!\!\!\!\!\!\! &\frac i{\sqrt 2},\!\!\!\!\!\!\!\! &0,\!\!\!\!\!\!\!\! &0,\!\!\!\!\!\!\!\! &0,\!\!\!\!\!\!\!\! &0,\!\!\!\!\!\!\!\! &0 \\
0,\!\!\!\!\!\!\!\! &0,\!\!\!\!\!\!\!\! &0,\!\!\!\!\!\!\!\! &0,\!\!\!\!\!\!\!\! &0,\!\!\!\!\!\!\!\! &0,\!\!\!\!\!\!\!\! &0,\!\!\!\!\!\!\!\! &0,\!\!\!\!\!\!\!\! &0,\!\!\!\!\!\!\!\! &0,\!\!\!\!\!\!\!\! &0,\!\!\!\!\!\!\!\! &0,\!\!\!\!\!\!\!\! &-\iem,\!\!\!\!\!\!\!\! &0,\!\!\!\!\!\!\!\! &0,\!\!\!\!\!\!\!\! &0,\!\!\!\!\!\!\!\! &0,\!\!\!\!\!\!\!\! &0,\!\!\!\!\!\!\!\! &0,\!\!\!\!\!\!\!\! &0,\!\!\!\!\!\!\!\! &0,\!\!\!\!\!\!\!\! &0,\!\!\!\!\!\!\!\! &0,\!\!\!\!\!\!\!\! &0,\!\!\!\!\!\!\!\! &0,\!\!\!\!\!\!\!\! &0,\!\!\!\!\!\!\!\! &0 \\
0,\!\!\!\!\!\!\!\! &0,\!\!\!\!\!\!\!\! &0,\!\!\!\!\!\!\!\! &0,\!\!\!\!\!\!\!\! &0,\!\!\!\!\!\!\!\! &0,\!\!\!\!\!\!\!\! &0,\!\!\!\!\!\!\!\! &0,\!\!\!\!\!\!\!\! &0,\!\!\!\!\!\!\!\! &0,\!\!\!\!\!\!\!\! &0,\!\!\!\!\!\!\!\! &0,\!\!\!\!\!\!\!\! &0,\!\!\!\!\!\!\!\! &0,\!\!\!\!\!\!\!\! &0,\!\!\!\!\!\!\!\! &0,\!\!\!\!\!\!\!\! &-\iem,\!\!\!\!\!\!\!\! &0,\!\!\!\!\!\!\!\! &0,\!\!\!\!\!\!\!\! &0,\!\!\!\!\!\!\!\! &0,\!\!\!\!\!\!\!\! &0,\!\!\!\!\!\!\!\! &0,\!\!\!\!\!\!\!\! &0,\!\!\!\!\!\!\!\! &0,\!\!\!\!\!\!\!\! &0,\!\!\!\!\!\!\!\! &0 \\
\end{array}
\rp
\feqn
\eqn
c_{75}=\lp
\begin{array}{ccccccccccccccccccccccccccc}
0,\!\!\!\!\!\!\!\! &0,\!\!\!\!\!\!\!\! &0,\!\!\!\!\!\!\!\! &0,\!\!\!\!\!\!\!\! &0,\!\!\!\!\!\!\!\! &0,\!\!\!\!\!\!\!\! &0,\!\!\!\!\!\!\!\! &0,\!\!\!\!\!\!\!\! &0,\!\!\!\!\!\!\!\! &0,\!\!\!\!\!\!\!\! &0,\!\!\!\!\!\!\!\! &0,\!\!\!\!\!\!\!\! &0,\!\!\!\!\!\!\!\! &0,\!\!\!\!\!\!\!\! &0,\!\!\!\!\!\!\!\! &0,\!\!\!\!\!\!\!\! &0,\!\!\!\!\!\!\!\! &0,\!\!\!\!\!\!\!\! &0,\!\!\!\!\!\!\!\! &0,\!\!\!\!\!\!\!\! &0,\!\!\!\!\!\!\!\! &0,\!\!\!\!\!\!\!\! &\frac i{\sqrt 2},\!\!\!\!\!\!\!\! &0,\!\!\!\!\!\!\!\! &0,\!\!\!\!\!\!\!\! &0,\!\!\!\!\!\!\!\! &0 \\
0,\!\!\!\!\!\!\!\! &0,\!\!\!\!\!\!\!\! &0,\!\!\!\!\!\!\!\! &0,\!\!\!\!\!\!\!\! &0,\!\!\!\!\!\!\!\! &0,\!\!\!\!\!\!\!\! &0,\!\!\!\!\!\!\!\! &0,\!\!\!\!\!\!\!\! &0,\!\!\!\!\!\!\!\! &0,\!\!\!\!\!\!\!\! &0,\!\!\!\!\!\!\!\! &0,\!\!\!\!\!\!\!\! &0,\!\!\!\!\!\!\!\! &0,\!\!\!\!\!\!\!\! &0,\!\!\!\!\!\!\!\! &0,\!\!\!\!\!\!\!\! &-\iem,\!\!\!\!\!\!\!\! &0,\!\!\!\!\!\!\!\! &0,\!\!\!\!\!\!\!\! &0,\!\!\!\!\!\!\!\! &0,\!\!\!\!\!\!\!\! &0,\!\!\!\!\!\!\!\! &0,\!\!\!\!\!\!\!\! &0,\!\!\!\!\!\!\!\! &0,\!\!\!\!\!\!\!\! &0,\!\!\!\!\!\!\!\! &0 \\
0,\!\!\!\!\!\!\!\! &0,\!\!\!\!\!\!\!\! &0,\!\!\!\!\!\!\!\! &0,\!\!\!\!\!\!\!\! &0,\!\!\!\!\!\!\!\! &0,\!\!\!\!\!\!\!\! &0,\!\!\!\!\!\!\!\! &0,\!\!\!\!\!\!\!\! &0,\!\!\!\!\!\!\!\! &0,\!\!\!\!\!\!\!\! &0,\!\!\!\!\!\!\!\! &0,\!\!\!\!\!\!\!\! &0,\!\!\!\!\!\!\!\! &0,\!\!\!\!\!\!\!\! &\iem,\!\!\!\!\!\!\!\! &0,\!\!\!\!\!\!\!\! &0,\!\!\!\!\!\!\!\! &0,\!\!\!\!\!\!\!\! &0,\!\!\!\!\!\!\!\! &0,\!\!\!\!\!\!\!\! &0,\!\!\!\!\!\!\!\! &0,\!\!\!\!\!\!\!\! &0,\!\!\!\!\!\!\!\! &0,\!\!\!\!\!\!\!\! &0,\!\!\!\!\!\!\!\! &0,\!\!\!\!\!\!\!\! &0 \\
\end{array}
\rp
\feqn
\eqn
c_{76}=\lp
\begin{array}{ccccccccccccccccccccccccccc}
0,\!\!\!\!\!\!\!\! &0,\!\!\!\!\!\!\!\! &0,\!\!\!\!\!\!\!\! &0,\!\!\!\!\!\!\!\! &0,\!\!\!\!\!\!\!\! &0,\!\!\!\!\!\!\!\! &0,\!\!\!\!\!\!\!\! &0,\!\!\!\!\!\!\!\! &0,\!\!\!\!\!\!\!\! &0,\!\!\!\!\!\!\!\! &0,\!\!\!\!\!\!\!\! &0,\!\!\!\!\!\!\!\! &0,\!\!\!\!\!\!\!\! &0,\!\!\!\!\!\!\!\! &0,\!\!\!\!\!\!\!\! &0,\!\!\!\!\!\!\!\! &0,\!\!\!\!\!\!\!\! &0,\!\!\!\!\!\!\!\! &0,\!\!\!\!\!\!\!\! &0,\!\!\!\!\!\!\!\! &0,\!\!\!\!\!\!\!\! &0,\!\!\!\!\!\!\!\! &0,\!\!\!\!\!\!\!\! &\frac i{\sqrt 2},\!\!\!\!\!\!\!\! &0,\!\!\!\!\!\!\!\! &0,\!\!\!\!\!\!\!\! &0 \\
0,\!\!\!\!\!\!\!\! &0,\!\!\!\!\!\!\!\! &0,\!\!\!\!\!\!\!\! &0,\!\!\!\!\!\!\!\! &0,\!\!\!\!\!\!\!\! &0,\!\!\!\!\!\!\!\! &0,\!\!\!\!\!\!\!\! &0,\!\!\!\!\!\!\!\! &0,\!\!\!\!\!\!\!\! &0,\!\!\!\!\!\!\!\! &0,\!\!\!\!\!\!\!\! &0,\!\!\!\!\!\!\!\! &0,\!\!\!\!\!\!\!\! &0,\!\!\!\!\!\!\!\! &-\iem,\!\!\!\!\!\!\!\! &0,\!\!\!\!\!\!\!\! &0,\!\!\!\!\!\!\!\! &0,\!\!\!\!\!\!\!\! &0,\!\!\!\!\!\!\!\! &0,\!\!\!\!\!\!\!\! &0,\!\!\!\!\!\!\!\! &0,\!\!\!\!\!\!\!\! &0,\!\!\!\!\!\!\!\! &0,\!\!\!\!\!\!\!\! &0,\!\!\!\!\!\!\!\! &0,\!\!\!\!\!\!\!\! &0 \\
0,\!\!\!\!\!\!\!\! &0,\!\!\!\!\!\!\!\! &0,\!\!\!\!\!\!\!\! &0,\!\!\!\!\!\!\!\! &0,\!\!\!\!\!\!\!\! &0,\!\!\!\!\!\!\!\! &0,\!\!\!\!\!\!\!\! &0,\!\!\!\!\!\!\!\! &0,\!\!\!\!\!\!\!\! &0,\!\!\!\!\!\!\!\! &0,\!\!\!\!\!\!\!\! &0,\!\!\!\!\!\!\!\! &0,\!\!\!\!\!\!\!\! &0,\!\!\!\!\!\!\!\! &0,\!\!\!\!\!\!\!\! &-\iem,\!\!\!\!\!\!\!\! &0,\!\!\!\!\!\!\!\! &0,\!\!\!\!\!\!\!\! &0,\!\!\!\!\!\!\!\! &0,\!\!\!\!\!\!\!\! &0,\!\!\!\!\!\!\!\! &0,\!\!\!\!\!\!\!\! &0,\!\!\!\!\!\!\!\! &0,\!\!\!\!\!\!\!\! &0,\!\!\!\!\!\!\!\! &0,\!\!\!\!\!\!\!\! &0 \\
\end{array}
\rp
\feqn
\eqn
c_{77}=\lp
\begin{array}{ccccccccccccccccccccccccccc}
0,\!\!\!\!\!\!\!\! &0,\!\!\!\!\!\!\!\! &0,\!\!\!\!\!\!\!\! &0,\!\!\!\!\!\!\!\! &0,\!\!\!\!\!\!\!\! &0,\!\!\!\!\!\!\!\! &0,\!\!\!\!\!\!\!\! &0,\!\!\!\!\!\!\!\! &0,\!\!\!\!\!\!\!\! &0,\!\!\!\!\!\!\!\! &0,\!\!\!\!\!\!\!\! &0,\!\!\!\!\!\!\!\! &0,\!\!\!\!\!\!\!\! &0,\!\!\!\!\!\!\!\! &0,\!\!\!\!\!\!\!\! &0,\!\!\!\!\!\!\!\! &0,\!\!\!\!\!\!\!\! &0,\!\!\!\!\!\!\!\! &0,\!\!\!\!\!\!\!\! &0,\!\!\!\!\!\!\!\! &0,\!\!\!\!\!\!\!\! &0,\!\!\!\!\!\!\!\! &0,\!\!\!\!\!\!\!\! &0,\!\!\!\!\!\!\!\! &\frac i{\sqrt 2},\!\!\!\!\!\!\!\! &0,\!\!\!\!\!\!\!\! &0 \\
0,\!\!\!\!\!\!\!\! &0,\!\!\!\!\!\!\!\! &0,\!\!\!\!\!\!\!\! &0,\!\!\!\!\!\!\!\! &0,\!\!\!\!\!\!\!\! &0,\!\!\!\!\!\!\!\! &0,\!\!\!\!\!\!\!\! &0,\!\!\!\!\!\!\!\! &0,\!\!\!\!\!\!\!\! &0,\!\!\!\!\!\!\!\! &0,\!\!\!\!\!\!\!\! &0,\!\!\!\!\!\!\!\! &0,\!\!\!\!\!\!\!\! &0,\!\!\!\!\!\!\!\! &0,\!\!\!\!\!\!\!\! &-\iem,\!\!\!\!\!\!\!\! &0,\!\!\!\!\!\!\!\! &0,\!\!\!\!\!\!\!\! &0,\!\!\!\!\!\!\!\! &0,\!\!\!\!\!\!\!\! &0,\!\!\!\!\!\!\!\! &0,\!\!\!\!\!\!\!\! &0,\!\!\!\!\!\!\!\! &0,\!\!\!\!\!\!\!\! &0,\!\!\!\!\!\!\!\! &0,\!\!\!\!\!\!\!\! &0 \\
0,\!\!\!\!\!\!\!\! &0,\!\!\!\!\!\!\!\! &0,\!\!\!\!\!\!\!\! &0,\!\!\!\!\!\!\!\! &0,\!\!\!\!\!\!\!\! &0,\!\!\!\!\!\!\!\! &0,\!\!\!\!\!\!\!\! &0,\!\!\!\!\!\!\!\! &0,\!\!\!\!\!\!\!\! &0,\!\!\!\!\!\!\!\! &0,\!\!\!\!\!\!\!\! &0,\!\!\!\!\!\!\!\! &0,\!\!\!\!\!\!\!\! &0,\!\!\!\!\!\!\!\! &\iem,\!\!\!\!\!\!\!\! &0,\!\!\!\!\!\!\!\! &0,\!\!\!\!\!\!\!\! &0,\!\!\!\!\!\!\!\! &0,\!\!\!\!\!\!\!\! &0,\!\!\!\!\!\!\!\! &0,\!\!\!\!\!\!\!\! &0,\!\!\!\!\!\!\!\! &0,\!\!\!\!\!\!\!\! &0,\!\!\!\!\!\!\!\! &0,\!\!\!\!\!\!\!\! &0,\!\!\!\!\!\!\!\! &0 \\
\end{array}
\rp
\feqn
\eqn
c_{78}=\lp
\begin{array}{ccccccccccccccccccccccccccc}
0,\!\!\!\!\!\!\!\! &0,\!\!\!\!\!\!\!\! &0,\!\!\!\!\!\!\!\! &0,\!\!\!\!\!\!\!\! &0,\!\!\!\!\!\!\!\! &0,\!\!\!\!\!\!\!\! &0,\!\!\!\!\!\!\!\! &0,\!\!\!\!\!\!\!\! &0,\!\!\!\!\!\!\!\! &0,\!\!\!\!\!\!\!\! &0,\!\!\!\!\!\!\!\! &0,\!\!\!\!\!\!\!\! &0,\!\!\!\!\!\!\!\! &0,\!\!\!\!\!\!\!\! &0,\!\!\!\!\!\!\!\! &0,\!\!\!\!\!\!\!\! &0,\!\!\!\!\!\!\!\! &0,\!\!\!\!\!\!\!\! &0,\!\!\!\!\!\!\!\! &0,\!\!\!\!\!\!\!\! &0,\!\!\!\!\!\!\!\! &0,\!\!\!\!\!\!\!\! &0,\!\!\!\!\!\!\!\! &0,\!\!\!\!\!\!\!\! &0,\!\!\!\!\!\!\!\! &\frac i{\sqrt 2},\!\!\!\!\!\!\!\! &0 \\
0,\!\!\!\!\!\!\!\! &0,\!\!\!\!\!\!\!\! &0,\!\!\!\!\!\!\!\! &0,\!\!\!\!\!\!\!\! &0,\!\!\!\!\!\!\!\! &0,\!\!\!\!\!\!\!\! &0,\!\!\!\!\!\!\!\! &0,\!\!\!\!\!\!\!\! &0,\!\!\!\!\!\!\!\! &0,\!\!\!\!\!\!\!\! &0,\!\!\!\!\!\!\!\! &0,\!\!\!\!\!\!\!\! &0,\!\!\!\!\!\!\!\! &0,\!\!\!\!\!\!\!\! &0,\!\!\!\!\!\!\!\! &0,\!\!\!\!\!\!\!\! &-\iem,\!\!\!\!\!\!\!\! &0,\!\!\!\!\!\!\!\! &0,\!\!\!\!\!\!\!\! &0,\!\!\!\!\!\!\!\! &0,\!\!\!\!\!\!\!\! &0,\!\!\!\!\!\!\!\! &0,\!\!\!\!\!\!\!\! &0,\!\!\!\!\!\!\!\! &0,\!\!\!\!\!\!\!\! &0,\!\!\!\!\!\!\!\! &0 \\
0,\!\!\!\!\!\!\!\! &0,\!\!\!\!\!\!\!\! &0,\!\!\!\!\!\!\!\! &0,\!\!\!\!\!\!\!\! &0,\!\!\!\!\!\!\!\! &0,\!\!\!\!\!\!\!\! &0,\!\!\!\!\!\!\!\! &0,\!\!\!\!\!\!\!\! &0,\!\!\!\!\!\!\!\! &0,\!\!\!\!\!\!\!\! &0,\!\!\!\!\!\!\!\! &0,\!\!\!\!\!\!\!\! &\iem,\!\!\!\!\!\!\!\! &0,\!\!\!\!\!\!\!\! &0,\!\!\!\!\!\!\!\! &0,\!\!\!\!\!\!\!\! &0,\!\!\!\!\!\!\!\! &0,\!\!\!\!\!\!\!\! &0,\!\!\!\!\!\!\!\! &0,\!\!\!\!\!\!\!\! &0,\!\!\!\!\!\!\!\! &0,\!\!\!\!\!\!\!\! &0,\!\!\!\!\!\!\!\! &0,\!\!\!\!\!\!\!\! &0,\!\!\!\!\!\!\!\! &0,\!\!\!\!\!\!\!\! &0 \\
\end{array}
\rp
\feqn
}

\end{appendix}

\end{document}